\documentclass[aps,pra,showpacs,amsmath,amssymb,lengthcheck]{revtex4}

\usepackage{color}
\usepackage{epsfig}

\begin{document}
\author{Bert Van Schaeybroeck}
\affiliation{Royal Meteorological Institute of Belgium, Ringlaan 3, BE-1180 Brussels, Belgium}
\author{Patrick Navez}
\affiliation{Department of Physics, Loughborough University, Loughborough, LE11 3TU, United Kingdom}
\author{Joseph O. Indekeu}
\affiliation{Institute for Theoretical Physics, KU Leuven, Celestijnenlaan 200 D, BE-3001 Leuven, Belgium}

\title{Interface potential and line tension for Bose-Einstein condensate mixtures\\ near a hard wall}
\date{\small\it \today}

\begin{abstract}
Within Gross-Pitaevskii (GP) theory we derive the interface potential $V(\ell)$ which describes the interaction between the interface separating two demixed Bose-condensed gases and an optical hard wall at a distance $\ell$. Previous work revealed that this interaction gives rise to extraordinary wetting
and prewetting phenomena. Calculations that explore non-equilibrium properties by using $\ell$ as a constraint provide a thorough explanation for this behavior. We find that at bulk two-phase coexistence, $V(\ell)$ for both complete wetting and partial wetting is monotonic with exponential decay. Remarkably, at the first-order wetting phase transition, $V(\ell)$ is independent of $\ell$. This anomaly explains the infinite continuous degeneracy of the grand potential reported earlier. As a physical application, using $V(\ell)$ we study the three-phase contact line where the interface meets the wall under a contact angle $\theta$. Employing an interface displacement model we calculate the structure of this inhomogeneity and its line tension $\tau$. Contrary to what happens at a usual first-order wetting transition in systems with short-range forces, $\tau$ does not approach a nonzero positive constant for $\theta \rightarrow 0$,  but instead approaches zero (from below) in the manner $\tau \propto -\theta$ as would be expected for a critical wetting transition. This hybrid character of $\tau$ is a consequence of the absence of a barrier in $V(\ell)$ at wetting. For a typical $V(\ell) = \mathcal{S} \exp (-\ell/\xi)$, with $\mathcal{S}$ the spreading coefficient, we conjecture that $\tau = -2 \,(1-\ln 2)\,\gamma \,\xi\,\sin\theta$ is exact within GP theory, with $\gamma$ the interfacial tension and $0 \leq \theta \leq \pi$.
\end{abstract}


\maketitle
\section{Introduction}
The experimental realization of Bose-Einstein condensation (BEC) in dilute Bose gases, now more than twenty-five years ago, initiated big experimental and theoretical advances in the field of ultracold gases~\cite{pita}. Most interesting in this context is that one gained immediate access to, and extended experimental control of the physics of, very pure quantum systems. For example, by means of  Feshbach resonances~\cite{inouye,stan,papp}, one is capable of tuning the interactions between the trapped atoms. Furthermore, an evanescent wave surface trap provides one with adjustable particle-wall interactions and permits one to approximate a ``hard wall" type of boundary ~\cite{rychtarik,perrin}. 

Uniform (flat-bottom) optical-box traps are now increasingly used to establish homogeneous ultracold gases in different dimensionality ~\cite{gaunt,navon}. This is particularly interesting from our perspective in this paper because a homogeneous system in semi-infinite geometry is the ideal theoretical setting for studying wetting phenomena. Therefore, results of experiments in flat-bottom traps can be compared straightforwardly with predictions of density-functional theories with (hard or soft wall) boundary conditions and without (harmonic) external potential.  Moreover, hybrid traps also exist that combine box-like confinement along two directions and harmonic along the third~\cite{mukherjee} whereby the local chemical potential can be adjusted. It has been suggested that these are of particular interest for studying interfaces~\cite{navon}.

Atomic BEC mixtures have been realized using either different isotopes or atomic species or by combining different hyperfine states of the same isotope. While weakly-demixed binary Bose-Einstein condensates (BECs)  were already observed more than 20 years ago~\cite{modugno,miesner,myatt,stamper2,hall,matthews}, strong phase separation was demonstrated only a decade later~\cite{mccarron,tojo,altin,papp2}. More recent experimental realizations of various new mixtures and their immiscibility properties include Cs and Yb~\cite{wilson}, $^{41}$K and $^{87}$Rb~\cite{burchianti}, $^{39}$K and $^{87}$Rb~\cite{lee}, $^{23}$Na and $^{87}$Rb~\cite{wang}.

In sum, the technological building blocks for experimentally investigating the physics of binary BECs near walls exist. Additionally, the experimental probing of an ultracold-gas interface was recently shown for a Bose-Fermi system~\cite{lous} while multiple studies stipulate the important role of interface physics in explaining equilibrium configurations of current-day experimental situations \cite{vanschaeybroeck2007, Grochowski,ruban,jimbo}. Interface statics \cite{IndekeuHanoi} and especially interface dynamics of multi-component condensates gained substantial recent attention~\cite{maity,balaz,indekeu2018,pal}.
\section{Wetting phase transition and interface potential} 

A wetting phase transition or, more generally, interface delocalization transition~\cite{cahn,ebner,nakanishi,binder} (for early reviews, see \cite{degennes,dietrich,bonn}), in its simplest form, takes place when one phase, say phase $1$, is expelled from the surface or ``wall" by another phase, $2$, which is then said to ``wet" the interface between the wall and phase $1$.  The wetting phase forms a macroscopic layer between the wall and phase $1$. The wetting transition corresponds to a singularity in the equilibrium surface excess (free) energy of the state in which phase $1$ is the phase present in bulk. This singularity manifests itself in the manner that Young's contact angle goes to zero when the wetting transition is approached from the partial wetting state or, simply, ``nonwet" state. 

When the equilibrium (excess) energy exhibits a discontinuity in its first derivative, the wetting transition is of first order. This is the case of concern in our present study. In a previous Letter the first-order wetting phase transition predicted by the GP theory for adsorbed binary mixtures of BECs at a hard wall was studied, as well as the accompanying prewetting phenomenon \cite{indekeu3}. Later work elaborated on this and studied also softer walls and critical wetting \cite{vanschaeybroeck2}. For the readers' convenience, in \cite{vanschaeybroeck2} a thorough description was provided of the set-up and principal formalism of the wetting phase transition in adsorbed BEC mixtures. In addition, a pedagogical introduction is available in Lecture Note form in \cite{indekeuFPSP}.

For our main purpose in this paper, being the derivation and application of an interface potential for adsorbed BECs, we would like to stress that some of the properties we will encounter possess close analogues in a mean-field type theory for another quantum system, the Ginzburg-Landau theory of superconductivity. The existence of (first-order and critical) interface delocalization transitions in surface-enhanced type-I superconductors was predicted in 1995~\cite{indekeu,indekeu2} and the later derivation of the interface potential for that system has been important to provide a deeper understanding and to offer further new physics (non-universal exponents for critical wetting, three-phase contact line structure)~\cite{blossey,boulter,vanleeuwen,vanleeuwen2,boulter2}. A succinct review including a summary of the experimental verification of wetting in superconductors can be found in~\cite{indekeuFPSP}.

The concept of an interface potential is a powerful tool for studying the wetting phase transition at a level which is more phenomenological (i.e., less microscopic) than a density-functional theory, but at the same time quantitatively precise for determining the character and associated singularities of phase transitions and critical phenomena. This is especially the case when the interface potential is used in an interface Hamiltonian theory in combination with a functional renormalization group approach \cite{bhl,lipowsky1,lipowsky2,lipowsky3,lipowsky4}. An early review \cite{dietrich} covers the uptake of this development and a later one reports on its subsequent advancement \cite{bonnRMP}. Recently, the usefulness and power of an interface-potential based approach was demonstrated in the ingenious non-local interface Hamiltonian theory of Parry and co-workers reviewed in \cite{parryswain,parryIH}.

The interface potential $V(\ell)$, in its simplest form, is a collective-coordinate representation of the excess (free) energy per unit area of a wetting film of prescribed thickness $\ell \,(\geq 0)$, regardless of whether or not this film is an equilibrium state of the system. The dependence on the phenomenological variable $\ell$ is obtained after integrating out the microscopic degrees of freedom and 
performing a partial partition sum over all configurations that satisfy the constraint of fixed $\ell$. The stable (or meta-stable) states are recovered at the global minimum  (or local minima) of this function, which can also be a boundary minimum (e.g., at $\ell =0$). However, the entire function $V(\ell)$ is useful when studying spatially inhomogeneous states which connect stable states via a path along which $\ell$ varies. An example of this is a drop or wedge configuration of an interface which meets the wall under a contact angle $\theta$, relevant to partial wetting states.

In this work we derive the interface potential for a Boson mixture
at zero temperature near a hard wall. Two phase-segregated species
are distinct in bulk and mutually permeate at their interface, the structure of which is
induced by three distinct atomic interactions. As compared to the
more usual one-component (liquid-gas or binary liquid mixture) interfacial system, this gives
rise to the existence of at least two additional length scales. We
present the interface potentials for five limiting regions of
parameter space using analytic methods. Although the system under
consideration concerns BEC phases, one may consider this work to
describe also the interface potential for a more general nonlinear two-component system.

As a new physical application, we employ $V(\ell)$ in an interface displacement model and calculate the structure and excess energy of an inhomogeneous state describing the three-phase contact line where the interface between the two condensates meets the optical wall under a contact angle $\theta$. The excess energy of this linear inhomogeneity is named the line tension and it has been the subject of exquisite curiosity and vigourous attention since, roughly, 1990. For a thorough discussion of line tension statistical mechanics, see \cite{SchimDiet}. Especially the behavior of the line tension upon approach of a wetting phase transition has been an arena  of lively debates and astonishing findings. For 
a review of line tension at wetting, see~\cite{indekeu4}. We will uncover that in this system of BECs adsorbed at a hard wall, in which the first-order wetting transition possesses extraordinary features, the line tension follows suit and displays a singularity at wetting that would normally be expected for critical wetting.

After introducing the set-up and the Gross-Pitaevskii formalism
in Sect.~\ref{MixtureofBosegases}, we recall the wetting phase
diagram for this set-up in Sect.~\ref{TheWettingDiagram} and
present the thermodynamics of a two-component interface potential
in Sect.~\ref{TheEffectiveInterfacePotential}. Its definition is given in Sect.~\ref{Definition} and 
Sect.~\ref{ADynamicalView} is devoted to a so-called ``dynamical
approach'' by means of which we illustrate our findings. A
discussion of the expected behavior of the interface potential is
given in Sect.~\ref{discussion}. Our results are then presented in
Sect.~\ref{bosonboson}. More specifically, in
Sects.~\ref{Asmallcoherencelength}
and~\ref{Alargecoherencelength}, we assume the healing length of
the adsorbed phase to be much longer than the healing length of
the bulk phase and vice versa while in
Sect.~\ref{LargeInterPhaseRepulsionK}, we deal with large
interspecies interactions or \textit{strong segregation}. Then, in
a fourth regime in Sect.~\ref{SmallxiandInfiniteK}, we introduce
and apply numerically a method to study the case of a strong
healing length asymmetry, combined with strong interspecies
repulsion. The case of \textit{weak segregation} and comparable
healing lengths of the phases, is studied
in~\ref{weaksegregationsection}. Based on the interface potential
we then calculate in a mean-field approach the structure of a three-phase contact line and its line tension in Sect.~\ref{linetension}. We conclude
in Sect.~\ref{interface potentialconclusion}.
The results we present are partly based on earlier unpublished work \cite{BVSPhD}. 

\section{Excess energy of Bose mixtures}\label{MixtureofBosegases}
Consider BEC gases $1$ and $2$, both at fixed chemical potentials
$\mu_{_2}$ and $\mu_{_1}$, respectively. Phase $2$ (when present) resides only in
the vicinity of the hard wall which is at $z=0$ whereas phase $1$
prevails far from the wall where it is the phase imposed in bulk. An additional translational symmetry
in the $x$-$y$ plane allows one to restrict attention to flat interfaces such
that the development of the interface potential becomes
essentially a one-dimensional problem. Weakly interacting BEC
gases at $T=0$ are well described by the ground state expectation
value of the boson field operator $\Psi_{_i}(z)$ with
$i=1,\,2$~\cite{fetter,pita}. In the absence of particle flow one
can choose the order parameters to be real valued such that the
excess grand potential per unit area can be cast in the form:
\begin{align}\label{grandpot}
\gamma(\mu_{_1},\mu_{_2})
&=\sum_{i=1,\,2}\left(\int_{z>0}{\text{d}z\,
\Psi_{_i}\left[-\frac{\hslash^{2}}{2m_{i}}
\boldsymbol{\nabla}^{2}-\mu_{_i}\right] \Psi_{_i}}\right.\nonumber\\
&\left.+\frac{G_{_{ii}}}{2}\Psi_{_i}^{4}\right)+\int_{z>0}{\text{d}
z\,\left[G_{_{12}}\Psi_{_1}^{2} \Psi_{_2}^{2}+P_{_1}\right]},
\end{align}
from which one derives the coupled time-independent
Gross-Pitaevskii (GP) equations by minimization with respect to
$\Psi_{_1}$ and $\Psi_{_2}$~\cite{indekeu3,vanschaeybroeck2}.
Interactions between atoms of species $i$ and $j$ are
characterized by the coupling constants $G_{_{ij}}=2\pi
\hslash^{2}a_{_{ij}}(m_{_i}^{-1}+m_{_j}^{-1})>0$ with $a_{_{ij}}$
the s-wave scattering lengths and $i,j=1,\,2$. Henceforth, we denote
the excess grand potential per unit area by the more convenient
term ``excess energy''.
 
\begin{figure}
    \epsfig{figure=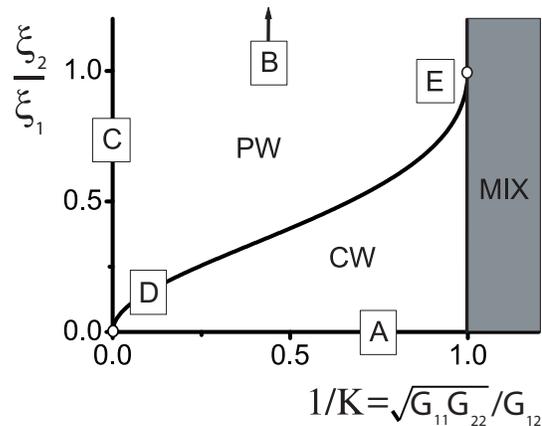,angle=0,width=200pt}
       \caption{The wetting phase diagram at two-phase coexistence ($\mu_{_2}/\overline{\mu}_{_2}=1$) with phase
       $1$ imposed far from the wall, as a function of $\xi_{_2}/\xi_{_1}$ and $1/K=\sqrt{G_{_{11}}G_{_{22}}}/G_{_{12}}$.
       The wetting line connects the points
       $(0,0)$ and $(1,1)$ and separates the regions of partial
       wetting (PW, no absorbed film of phase 2) and complete wetting (CW, a macroscopic wetting layer of phase 2). When $1/K>1$, phases $1$ and $2$ mix and there, a (metastable)
       $1$-$2$ interface does not exist. The region indicated by letter A is treated in Sect.~\ref{Asmallcoherencelength}, region B with $\xi_{_2}/\xi_{_1}\rightarrow \infty$ in
    Sect.~\ref{Alargecoherencelength} and analogously for regions C, D and E.
       \label{fig0}
       }
\end{figure} 

The imposed boundary conditions are:
\begin{align}\label{bound}
\Psi_{_2}(z=0)=\Psi_{_1}(0)=0,\,\Psi_{_2}(\infty)=0\text{ and }
\Psi_{_1}(\infty)\equiv \sqrt{n_{_1}},
\end{align}
where $n_{_1}$ is the number density of the pure phase of condensate $1$
with fixed chemical potential $\mu_{_1}$ and self-interaction
$G_{_{11}}$, i.e., $n_{_1} = \mu_{_1}/G_{_{11}}$. Note that the particle-wall interactions are solely
mediated by the first two conditions in \eqref{bound}. The pure bulk phase pressures $P_{_i}$ and the chemical
potential $\mu_{_i}$ are related by $P_{_i}=\mu_{_i}^2/(2G_{_{ii}})$ with
$i=1,\,2$. Therefore, each value of $\mu_{_1}$ has an associated
chemical potential for phase $2$, defined by
$\overline{\mu}_{_2}\equiv\mu_{_1}\sqrt{G_{_{22}}/G_{_{11}}}$,
such that at two-phase coexistence, i.e., when $P_{_2}=P_{_1}$,
$\mu_{_2}$ equals $\overline{\mu}_{_2}$. We define $\overline{n}_2 \equiv \overline{\mu}_{_2}/G_{_{22}}$.

\section{The Wetting Phase Diagram}\label{TheWettingDiagram}
For the pure phases $1$ and $2$, the typical lengths of variation
of wave functions $\Psi_{_1}$ and $\Psi_{_2}$ are the healing
lengths $\xi_{_1}\equiv\hslash/\sqrt{2m_{_1}\mu_{_1}}$ and
$\xi_{_2}\equiv\hslash/\sqrt{2m_{_2}\overline{\mu}_{_2}}$,
respectively. We introduce two bulk parameters, namely
$\xi_{_2}/\xi_{_1}$ and the inter-phase interaction parameter
$K\equiv G_{_{12}}/\sqrt{G_{_{11}}G_{_{22}}}$. One can then
rewrite $K$ and $\xi_{_2}/\xi_{_1}$ as a function of the masses
$m_{_i}$ and scattering lengths $a_{_{ij}}$ as
follows~\cite{pita}:
\begin{align}\label{kenxi}
K=\frac{m_{_1}+m_{_2}}{2\sqrt{m_{_1}m_{_2}}}\frac{a_{_{12}}}{\sqrt{a_{_{11}}a_
{_{22}}}}\text{
 and  }\xi_{_2}/\xi_{_1}=\sqrt[4]{\frac{m_{_1}a_{_{11}}}{m_{_2}a_{_{22}}}}.
\end{align}
For our purposes we assume $P_{_1} \geq P_{_2}$. In order to obtain pure phase 1 as the stable phase in bulk, $K$ must
be larger than $\mu_{_2}/\overline{\mu}_{_2}=\sqrt{P_{_2}/P_{_1}}$,
where the external parameter $1-\mu_{_2}/\overline{\mu}_{_2}$
quantifies the deviation from bulk two-phase coexistence.  If, in addition, $P_{_1}=P_{_2}$, pure phase 1 and pure phase 2 coexist in bulk and the condition for phase separation then becomes $K>1$~\cite{ao}.

The wetting phase diagram at bulk two-phase coexistence is depicted in
Fig.~\ref{fig0}. It shows that complete wetting is possible for practically every
value of $K$ whenever $\xi_{_2}\ll\xi_{_1}$ and it can also occur for practically every value of $\xi_{_2}/\xi_{_1} \,(<1)$ in the regime of weak segregation, i.e., $(0<)\, K-1 \ll 1$. The wetting
line (WL), separating the partial wetting (PW) and the complete
wetting (CW) regions indicates a first-order surface phase transition  \cite{indekeu3}. On
the WL, arbitrary film thicknesses can occur and are associated with an infinite degeneracy of
the grand potential. This entails another peculiarity off of
coexistence where this first-order WL is continued by a nucleation
or prewetting line which is all the way of second order and, contrary
to expectations, not tangential to the
line of bulk coexistence at the wetting transition point~\cite{indekeu3}.

The prewetting surface, which contains the WL at
$\mu_{_2}/\overline{\mu}_{_2}=1$,
satisfies~\cite{vanschaeybroeck2}:
\begin{align}\label{prewett}
\sqrt{K-\mu_{_2}/\overline{\mu}_{_2}}=
\frac{\sqrt{2}}{3}\left(\frac{\mu_{_2}/\overline{\mu}_{_2}}{\xi_{_2}/\xi_{_1}}-\xi_{_2}/\xi_{_1}\right).
\end{align}
In Sect.~\ref{bosonboson}, we investigate the regions in the
$(K,\xi_{_2}/\xi_{_1})$-space, indicated in Fig.~\ref{fig0} by the
letters A through E.
\begin{figure}
\begin{center}
   \epsfig{figure=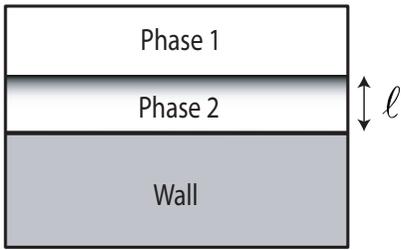,angle=0,width=150pt}
       \caption{Configuration with pure phase 1 stable in bulk and an adsorbed film of pure species 2 at the wall. This configuration is homogeneous (i.e., translationally invariant in directions parallel to the wall). The interface potential $V(\ell)$ maps a configuration with a microscopic adsorbed film of species $2$ of thickness
    $\ell$, or a macroscopic wetting layer of pure phase 2 ($\ell \rightarrow \infty$), to its excess grand potential.}
       \label{fig1}
       
       \end{center}
\end{figure}
\section{The Interface Potential}\label{TheEffectiveInterfacePotential}
\subsection{Definition}\label{Definition}
The interface potential $V(\ell)$ relates a system with phase 1 imposed in bulk and
an adsorbed film of phase 2 of thickness $\ell$ to its excess energy per unit
area. In addition, in our convention the potential vanishes for a configuration with
an infinitely thick film (macroscopic layer) at coexistence with the bulk. Therefore,
one can write:
\begin{equation}
\label{intpotdef}
V(\ell)\equiv \gamma(\mu_{_1},\mu_{_2};\ell) -\gamma_{_0},
 \end{equation}
where we have added the argument $\ell$ in the functional $\gamma$ to highlight that $\gamma$ is evaluated for an adsorbed film of imposed thickness $\ell$, and where
\begin{equation}
\gamma_{_0}\equiv\left.\gamma(\mu_{_1},\overline{\mu}_{_2};\ell)\right|_{\ell=\infty}
\end{equation}
We
define the \textit{film thickness $\ell$} as:
\begin{align}\label{ell}
\ell\equiv\Gamma/\overline{n}_{_2}\equiv 
\int_{_0}^{_\infty}\text{d}z\,\psi_{_2}^2 ,
\end{align}
where $\Gamma$ is the adsorption of species $2$, being the excess
particle number of species $2$ per unit area. This definition is a physically obvious way of relating the length $\ell$ to the (in principle) measurable quantity $\Gamma$. It is fortunate that the second power of $\psi$ is featured in the integral constraint \eqref{ell}. This implies that the constrained profiles are analytic everywhere (except at boundaries) and that the singularities that are known to occur when a crossing criterion is used \cite{boulter2} are absent in our approach.

Generally, states with arbitrary film thickness $\ell$ are
unstable as they do not obey the principle of minimal energy for
$\gamma$. The standard procedure to derive them nevertheless, using a variational principle, is to
constrain state space to states with film thicknesses $\ell$,
\textit{while} assuming the potential $\gamma$  still describes
the energy of the system. Thus, instead of $\gamma$, one must
minimize
\begin{align}
\widehat{\gamma}(\mu_{_1},\mu_{_2};\Pi)=\gamma(\mu_{_1},\mu_{_2};\ell)+\Pi
\ell,
\end{align}
with respect $\psi_{_1}$ and $\psi_{_2}$. Here the disjoining
pressure
\begin{align}
\Pi\equiv -\partial V(\ell)/\partial \ell,
\end{align}
 plays the role of a
Lagrange multiplier. One therefore proceeds by first performing a
variational procedure in order to get the optimal film thickness
$\ell$ at fixed disjoining pressure, and subsequently deducing
the interface potential. We denote the equations of state which
minimize $\widehat{\gamma}$ with respect to the bosonic wave
functions, by the \textit{modified GP equations}. When the
disjoining pressure vanishes, one recovers the equilibrium
solutions for $\gamma$ which have the property to have no force
exerted on the $1$-$2$ interface. In addition, one must also allow for the existence of boundary minima at which $\Pi \neq 0$, notably at $\ell =0$. Global minima in the form of boundary minima may also correspond to equilibrium solutions.  

\subsection{A Dynamical Point of View}\label{ADynamicalView}
By a ``mechanical" analogy, used throughout this work, the
modified GP equations can be reinterpreted as Newton's equations of
motion for two particles where the evolution in time must be
replaced by a variation of the space coordinate $z$. The particles
have one-dimensional ``positions'' $\psi_{_i}$ and ``masses''
$\hslash^2/(2m_{_i}\mu_{_i})$ and they move with kinetic energy
$T$ in a potential $U$ (both per unit volume)~\cite{footnote1}.
Then,
\begin{subequations}
\begin{align}\label{energies}
U[\psi_{_1},\psi_{_2}]&=2P_{_1}\left[\psi_{_1}^{2}+\eta\psi_{_2}^{2}-
\frac{\psi_{_1}^{4}}{2}-\frac{\psi_{_2}^{4}}{2}-K
\psi_{_1}^{2}\psi_{_2}^{2}\right],\\
T[\dot{\psi}_{_1},\dot{\psi}_{_2} ]&=2P_{_1}\left[
\xi_{_1}^2\dot{\psi^2_{_1}}+\xi_{_2}^2\dot{\psi^2_{_2}}\right].
\end{align}
\end{subequations}
Here the dot denotes the derivative with respect to $z$,
 $\psi_{_1}\equiv\Psi_{_1}/\sqrt{n_{_1}}$ and
$\psi_{_2}\equiv\Psi_{_2}/\sqrt{\overline{n}_{_2}}$.
Here, the dimensionless parameter
\begin{equation}
\label{etadef}
\eta\equiv\frac{\mu_{_2}}{\overline{\mu}_{_2}}-\frac{\Pi}{2P_{_1}},
\end{equation}
takes over the role of Lagrange multiplier from $\Pi$.
Importantly, there is a one-to-one correspondence between the wave function
profiles of thickness $\ell$ and the parameter $\eta$; therefore
all wave profiles for configurations with fixed film thickness
$\ell$ are the same, independently of
$\mu_{_2}/\overline{\mu}_{_2}$. In other words, the nonequilibrium and equilibrium profiles of equal thickness $\ell$ are analytic and are the same at two-phase coexistence and off of coexistence, respectively, provided the latter exist as equilibrium states. A conservation of energy per unit
volume links the kinetic to the potential energy at each
point:
\begin{align}\label{consenergy}
U[\psi_{_1},\psi_{_2}]+T[\dot{\psi}_{_1},\dot{\psi}_{_2}]=P_{_1},
\end{align}
obtained by a summation of the first integrals of the modified GP
equations~\cite{footnote2}.

\subsection{Discussion on $V(\ell)$}\label{discussion}
We discuss the behavior of the interface potential before its
explicit calculation. First, introduce the surface tensions of the pure phases 1 and 2 when adsorbed at the wall,
$\gamma_{_{1W}}\equiv\gamma(\mu_{_1},\overline{\mu}_{_2};\ell =0)$ and
$\gamma_{_{2W}}\equiv\gamma(\mu_{_1},\overline{\mu}_{_2};\ell =\infty) - \gamma_{_{12}}$,
 \textit{defined at two-phase coexistence} using only $\mu_{_1}$
and $\overline{\mu}_{_2}$ and not $\mu_{_2}$, and with $\gamma_{_{12}}\equiv\gamma_{_
{12}}(\mu_{_1},\overline{\mu}_{_2})$ the tension of the $1$-$2$ interface. Expressions for
$\gamma_{_{12}}$ can be found in~\cite{vanschaeybroeck}
and references therein; its qualitative dependence on $K$ is
rather simple: $\gamma_{_{12}}$ is maximal in the limit of
$K\rightarrow \infty$, decreases upon decrease of $K$ and vanishes
at $K=1$.

It is instructive to reexpress the interface potential by use of
\eqref{consenergy} in the following form:
\begin{align}\label{simpleform2}
V(\ell)+\gamma_{_0}
&=2P_{_1}(1-\mu_{_2}/\overline{\mu}_{_2})\ell + 2P_{_1}(\eta-1)\ell\\
&+2\int_{_0}^{_\infty}\text{d}z\, T[\dot{\psi}_{_1}(z),\dot{\psi}_{_2}(z)]
\nonumber.
\end{align}
Remarkably, in~\eqref{simpleform2} we were able to isolate the
part dependent on the deviation from bulk coexistence, which is
the term proportional to $1-\mu_{_2}/\overline{\mu}_{_2}$. Indeed,
as mentioned before, the wave functions for films with thickness
$\ell$ are the same for different values of
$1-\mu_{_2}/\overline{\mu}_{_2}$.

Starting with a configuration with only species $1$ at the wall,
one gets
$\left.V(\ell)\right|_{\ell=0}=\gamma_{_{1W}}-\gamma_{_0}$
and since
$\left.V(\ell)\right|_{\ell=\infty} + \gamma_{_0} =\gamma_{_{2W}}+\gamma_{_{12}}$,
we have that the difference $\left.V(\ell)\right|_{\ell=0}- \left.V(\ell)\right|_{\ell=\infty}$ corresponds to
the \textit{spreading coefficient} $\mathcal{S}$\label{spreaddef}:
\begin{align}
\mathcal{S}\equiv\gamma_{_{1W}}-(\gamma_{_{2W}}+\gamma_{_{12}}).
\end{align}
For a system at two-phase coexistence and in equilibrium, the spreading coefficient $\mathcal{S}$ is negative (for partial wetting) or zero (for complete wetting). A positive $\mathcal{S}$ corresponds to a nonequilibrium state which, in the course of time, may relax to a complete wetting equilibrium state (of lower excess energy) with $\mathcal{S}=0$. Note that since we define the interface potential to vanish in the limit $\ell=\infty$ at bulk two-phase coexistence, the partial wetting state ($\ell =0$ in our system) satisfies $\left.V(\ell)\right|_{\ell=0} = \mathcal{S}$.

To understand qualitatively how the parameters $\xi_{_2}/\xi_{_1}$
and $K$ influence the interface potential, we consider first a simple
\textit{one-component} (e.g., an adsorbed liquid-vapor or Ising-like) system. There, the form
of the interface potential in the presence of interactions of
short-range nature (ignoring van der Waals forces), is:
\begin{align}\label{FJexpr}
V(\ell)=h \ell +\sum_{m=1}^\infty
A_{_m}e^{-m\ell/\xi_{_c}}.
\end{align}
Here $h$ is the bulk field and the expansion coefficients $A_{_m}$
are independent of $\mu_{_2}/\overline{\mu}_{_2}$. The length
$\xi_{_{c}}$ corresponds to the bulk correlation length which is
also the decay length of the tail in the interface
profile. The dominant variation for large $\ell$ comes
from the first term $A_{_{1}}e^{-\ell/\xi_{c}}$, at least when
$A_{_{1}}\neq 0$. 

Further on we will establish that an expression akin to \eqref{FJexpr}, with exponentially decaying terms, gives the
generic interface potential for the two-component BEC system. From \eqref{simpleform2}, one sees that the bulk
field corresponds to:
\begin{align}\label{bulkfield}
h=2P_{_1}(1-\mu_{_2}/\overline{\mu}_{_2}).
\end{align}
In analogy with the one-component system, one might expect the length
$\xi_{_{c}}$ in the binary system to be either $\xi_{_1}$ or
$\xi_{_2}$. However, we find that this is only true when the
mutual penetration is small. Generally, $\xi_{_{c}}$ also depends
on the interspecies repulsion parameter $K$.

Yet, \eqref{FJexpr} turns out inadequate to describe the
interface potential in two regimes: 1) in
Sect.~\ref{Asmallcoherencelength}, long-range correlations appear
and $V(\ell)$ displays an extraordinary algebraic decay $V(\ell)\propto \ell^{-1}$ for large $\ell$, and 2)
instead of one, two length scales determine the exponential decay
of the interface potential found in
Sect.~\ref{SmallxiandInfiniteK}. The competition and crossover
behavior in the presence of two characteristic lengths was earlier
encountered in~\cite{hauge2} and in~\cite{vanleeuwen,KogaIW} where it gave rise to
non-universal critical wetting exponents that depend on the ratio of these two lengths.

One may wonder whether~\eqref{bulkfield} is indeed what one
expects for the bulk field. For a (nonequilibrium) system with a large film
thickness $\ell$, the exponential contributions in~\eqref{FJexpr}
vanish such that the excess energy should be $\ell(P_{_1}-P_{_2})$
with $P_{_i}$ minus the derivative of $\Omega=\gamma-P_{_1}\mathcal{V}$ with
respect to the volume $\mathcal{V}$, evaluated for pure bulk phase $i$.
However, the bulk field in~\eqref{bulkfield} is not exactly given by $P_{_1}-P_{_2}$, for the following reasons.
The method to construct the state with large film thickness
$\ell$, is to minimize the $\widehat{\gamma}$ which is constrained
because of the applied disjoining pressure. This results in an
equal pressure for pure phase $1$ and pure phase $2$, where the
pressure is now the derivative of
$\widehat{\Omega}=\widehat{\gamma}-P_{_1}\mathcal{V}$ with respect to the
volume, again evaluated for the pure bulk phase. One easily
calculates that the pure bulk phase density of phase $2$, obtained
with $\widehat{\Omega}$, equals $\overline{\mu}_{_2}/G_{_{22}}$,
rather than the density $\mu_{_2}/G_{_{22}}$ as obtained with
$\Omega$~\cite{footnote3}. 
Eventually,
$h=P_{_1}[-2(\mu_{_2}/\overline{\mu}_{_2})+1]-(-P_{_1})$ which is
~\eqref{bulkfield}. Note that a different choice of definition
for $\ell$ would yield a different $h$.
\begin{figure}
\begin{center}
    \epsfig{figure=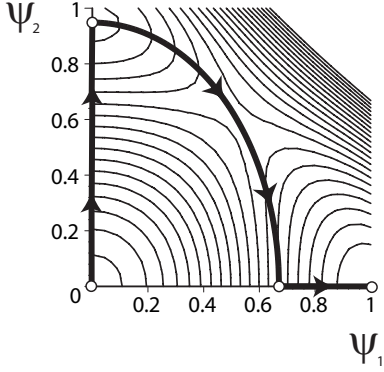,angle=0,width=200pt}
       \caption{The trajectory in the coordinates $(\psi_{_1},\psi_{_2})$
       for $\xi_{_2}/\xi_{_1}=0$, $K=2$ ($\eta=0.9$) with
       equipotential curves of $U$ in the background. The points $(1,0)$ and $(0,1)$ are maxima
       of $U$, whereas $(0,0)$ is a minimum and there exists a saddle point for the mixed phase in the middle.
       The upper right equipotential curves were left out due to their large
       density.\label{fig2a}
        }
\end{center}
\end{figure}
\section{Results for the interface potential}\label{bosonboson}
\subsection{Strong Healing Length Asymmetry I}\label{Asmallcoherencelength}
We focus in the following on the situation in which the adsorbed
species $2$ has a much smaller healing length than species $1$ or
$\xi_{_2}\ll \xi_{_1}$. Looking at region 1 in Fig.~\ref{fig0},
complete wetting is expected at coexistence since $\mathcal{S}$ is
identically zero. Indeed, first of all $\gamma_{_{2W}}$ is much
smaller than the metastable extension of $\gamma_{_{1W}}$, denoted
by $\gamma_{_{1W}}^*$ because $\gamma_{_{2W}}\propto\xi_{_2}$ and
$\gamma_{_{1W}}^*\propto\xi_{_1}$, and secondly,
$\gamma_{_{12}}\le\gamma_{_{1W}}$. This ascertains the zero
spreading coefficient.

According to the dynamical two-particle approach,
$\xi_{_2}/\xi_{_1}=0$ means that particle 2 cannot gain momentum
and will adapt its position to the potential which is modified by
the moving particle $1$. We plot the evolution of the particle
positions $(\psi_{_1},\psi_{_2})$ in Fig.~\ref{fig2a} for a system
with nonzero film thickness and $\xi_{_2}/\xi_{_1}=0$. Starting in
the point $(0,0)$, particle $2$ jumps to position $\sqrt{\eta}$ on
the short time scale $\xi_{_2}$. Then, particle $1$ starts to
convert its potential to kinetic energy on a time scale
$\Lambda_{_1}=\xi_{_1}/\sqrt{K-1}$, as seen from the modified GP
equations since $\eta\rightarrow 1$ when $\ell\rightarrow \infty$:
\begin{align}\label{curve}
\xi_{_1}^2\ddot{\psi}_{_1}=\psi_{_1}\left[\eta
K-1+\psi_{_1}^2(1-K^2)\right]\,\text{ and }\,
\psi_{_2}^2=\eta-K\psi_{_1}^2.
\end{align}
Then, particle 2 is stopped abruptly at the position $\psi_{_2}=0$
when $\psi_{_1}=\sqrt{\eta/K}$. This abruptness is allowed by a
lack of momentum of the massless particle $2$. Subsequently,
particle $1$ continues climbing the potential hill, reaching the
top $\psi_{_1}=1$ only after an infinite amount of time.
\begin{figure}
\begin{center}
    \epsfig{figure=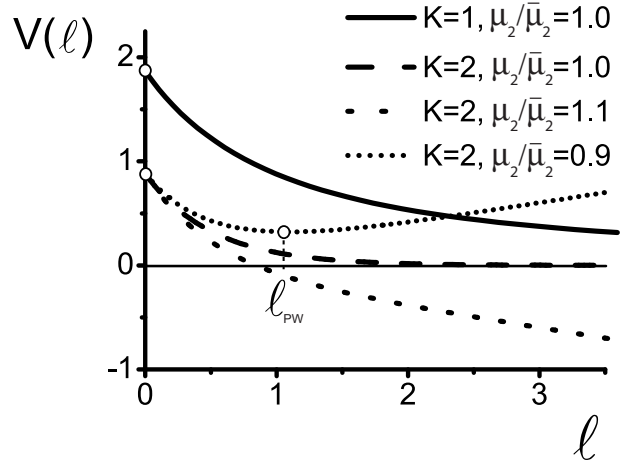,angle=0,width=230pt}
       \caption{
       Interface potential $V(\ell)$ in units of $P_{_1}\xi_{_1}$ as a function of the film thickness $\ell$ in units of
       $\xi_{_1}$ when $\xi_{_2}/\xi_{_1}=0$. The curve for $K=2$
       and $\mu_{_2}/\overline{\mu}_{_2}=1$ follows an exponential decay whereas the slow
       decay for $K=1$ is described by a power law. The thickness $\ell_{_{\rm PW}}$ indicates
       the thickness of the prewetting film for $K=2$ and $\mu_{_2}/\overline{\mu}_{_2}=0.9$ when phase 1 is stable in bulk and phase 2 is metastable. No such local minimum, or prewetting film of phase 2, exists when $\mu_{_2}/\overline{\mu}_{_2} >1$ because phase 1 is not stable in bulk.\label{fig3}
}
\end{center}
\end{figure}
The particles then follow the equations of motion:
\begin{align}\label{horizontal1}
\xi_{_1}^2\ddot{\psi}_{_1}=\psi_{_1}\left[-1
+\psi_{_1}^2\right]\quad\text{ and }\quad \psi_{_2}=0.
\end{align}

In the Appendix we give the exact interface potential
$V(\ell)$ together with the applied analytic methods for
this case $\xi_{_2}/\xi_{_1}=0$. We show $V(\ell)$ in
Fig.~\ref{fig3} for $K=1$ and $K=2$, the latter for different
values of $\mu_{_2}/\overline{\mu}_{_2}$. For large $\ell$ and
$K\neq 1$, we derive the leading terms:
\begin{align}\label{nogesnenandere}
V(\ell)=h \ell +\overline{A}_{_{1}}P_{_1} \xi_{_1}
e^{-2\ell\sqrt{K+1}/\xi_{_1}}+\ldots
\end{align}
with $\overline{A}_{_{1}}$ given in~\eqref{weennul} and
$\overline{A}_{_{1}}\rightarrow 0$ when $K\rightarrow1$ and
$K\rightarrow\infty$. The fact that the decay length
$\xi_{_c}=\xi_{_1}/(2\sqrt{K+1})$ contains the parameter $K$
expresses that the film thickness is chiefly modified by a
changing overlap between both species. The interface potential is
a monotonically decaying function for which both the decay length
$\xi_{_c}$ and $V(0)$ decrease with increasing $K$ (see
Fig.~\ref{fig3}). Indeed, upon increase of $K$, $V(0)$ goes
down due to the increase of $\gamma_{_{12}}$.

For low pressure of the adsorbed species $2$, when
$\mu_{_2}/\overline{\mu}_{_2}<1$, the energy necessary to adsorb a
large film increases linearly with its thickness. Nevertheless,
for every $\mu_{_2}/\overline{\mu}_{_2}$, there exists an energy
minimum for thin prewetting film thickness $\ell_{_{\rm PW}}$ (short
dotted line in Fig.~\ref{fig3}). By a simple calculation using
\eqref{nogesnenandere}, it is shown that the prewetting film
thickness $\ell_{_{\rm PW}}$ diverges logarithmically slowly for 
$\mu_{_2}/\overline{\mu}_{_2} \uparrow 1$. Its asymptotic behavior is given by
\begin{equation}
\ell_{_{\rm PW}}\sim-(\xi_{_1}/(2\sqrt{K+1}))\ln(1-\mu_{_2}/\overline{\mu}_{_2}).
\end{equation}

As weak segregation is approached, i.e., $K\rightarrow 1$,
the length $\xi_{_c}\rightarrow \xi_{_1}/(2\sqrt{2})$ remains
finite, while the coefficient $\overline{A}_{_{1}}$
vanishes \cite{footnote4}. At the bulk demixing point, i.e., when
$K=1$, one can obtain from \eqref{intpot1} a power-law form
for the interface potential:
\begin{align}\label{power}
V(\ell)=h\ell +
\frac{\pi^2P_{_1}\xi_{_1}}{8(\ell/\xi_{_1})}-\frac{\pi^{3}P_{_1}\xi_{_1}}{256(\ell/\xi_{_1})^{3}}
+\ldots
\end{align}
This potential is also depicted in Fig.~\ref{fig3} for
$\mu_{_2}/\overline{\mu}_{_2}=1$. From \eqref{power}, one can
notice the
long-ranged nature (algebraic decay) of $V(\ell)-h\ell $ since it is proportional to $1/\ell$.  This interface potential predicts that the equilibrium configuration in the limit $K \downarrow 1$ consists of a 1-2 interface that is delocalised from the surface. This interface, at $K=1$, has interfacial tension zero but nevertheless possesses a non-vanishing wave function profile that connects the two pure phases in bulk and that is of infinite width as determined by the diverging interspecies penetration depths $\xi_i/\sqrt{K-1}$, $i=1,2$. This divergence is reminiscent of a divergent correlation length and in that sense the limit $K \downarrow 1$ is akin to an approach to criticality, {\em but with still two distinct coexisting pure phases 1 and 2 in bulk}. 
\begin{figure}
\begin{center}
        \epsfig{figure=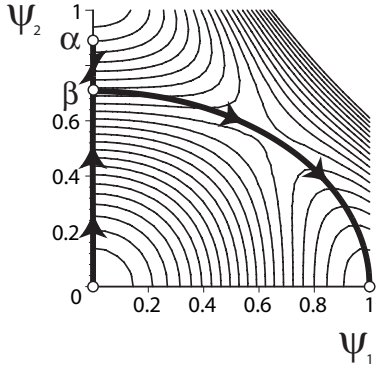,angle=0,width=140pt}
       \caption{The trajectory for $(\psi_{_1},\psi_{_2})$ with $\xi_{_1}/\xi_{_2}=0$, $K=2$ and
       $\eta=1.02$. Starting at position $(0,0)$, the path goes to
       $\alpha=(\sqrt{\eta},0)$ where it turns back to
       the point $\beta=(\sqrt{1/K},0)$. Further, both
       $\psi_{_1}$ and $\psi_{_2}$ are nonzero up to the point
       $(0,1)$.\label{fig4}
       }
\end{center}
\end{figure}\begin{figure}
\begin{center}
    \epsfig{figure=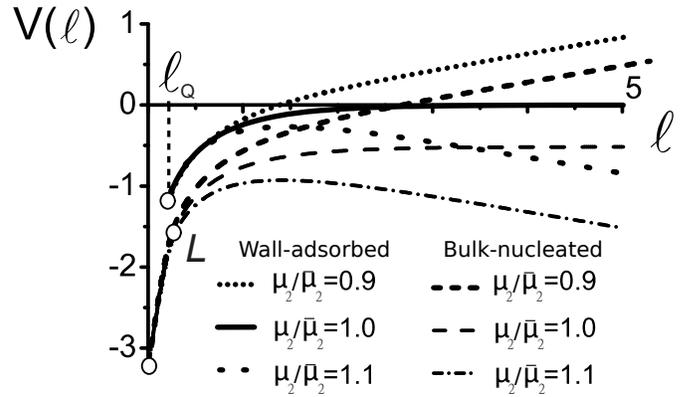,angle=0,width=250pt}
       \caption{Interface potential $V(\ell)$ in units of $P_{_1}\xi_{_2}$ as a function of the film thickness $\ell$ in units of
       $\xi_{_2}$ for the wall-adsorbed and
       the bulk-nucleated states when $\xi_{_1}/\xi_{_2}=0$ and $K=5$ and for different
       $\mu_{_2}/\overline{\mu}_{_2}$. The point $L$ indicates a stratification point where
       pure phase $1$ is separated into two regions due to a planar
       film
       of pure phase $2$ parallel to but infinitely far from the wall. Note that the thickness $\ell$ which corresponds to the
       point $L$ does not equal $\ell_{_Q}$. The latter is the initial point of the
       potential of the wall-adsorbed state for all values of $\mu_{_2}/\overline{\mu}_{_2}$. However,
       the values of the potential for the three different values of $\mu_{_2}/\overline{\mu}_{_2}$ at $\ell_{_Q}$, differ
       slightly; this is not visible. It is obvious that we are in the partial wetting regime
       and that for all values for $\ell$, the bulk-nucleated state has the lower excess energy. \label{fig5}
       }
\end{center}
\end{figure}

The previous discussion was for $\xi_{_2}/\xi_{_1} =0$. If one sets $\xi_{_2}/\xi_{_1}$ to a small but non-zero value, the
most significant change to $V(\ell)$ is a decrease of the spreading coefficient $V(0)$, caused by a modification of the wave function $\psi_{_2}$ at both locations where it vanishes~\cite{vanschaeybroeck}. One location is near the surface, where a negative correction linear in $\xi_{_2}/\xi_{_1}$ is incurred. The second location is in the 1-2 interface where a positive correction quadratic in $\xi_{_2}/\xi_{_1}$ results \cite{vanschaeybroeck}.

\subsection{Strong Healing Length Asymmetry II}\label{Alargecoherencelength}
We concentrate now on  the inverse case $\xi_{_1}\ll\xi_{_2}$. As
one expects, partial wetting is met, evidenced by a negative
spreading coefficient since $\gamma_{_{1W}}\ll\gamma_{_{2W}}$.
Species $2$ is so strongly disfavored near the wall, that it
rather nucleates in the bulk. Indeed, we find that ``wall-adsorbed
states'' attain a higher energy than ``bulk-nucleated states''
with the same thickness $\ell$ and where the latter are planar
bulk states or essentially one-dimensional stationary soliton
states. Henceforth, we focus on the limiting case $\xi_{_1}/\xi_{_2}=0$. A general two-particle trajectory for wall-adsorbed states
is given in Fig.~\ref{fig4}. Particle $2$ starts at position $(0,\,0)$
and reaches its maximal position $\alpha$ over a time scale
$\xi_{_2}$, where it reverts its motion, up to the point
$\beta=(0,\sqrt{1/K})$. This evolution is described by:
\begin{align}\label{horizontal2}
\xi_{_2}^2\ddot{\psi}_{_2}=\psi_{_2}\left[-\eta
+\psi_{_2}^2\right]\quad\text{ and }\quad \psi_{_1}=0.
\end{align}
When $\psi_{_2}$ reaches the point $\beta$ for the second time,
$\psi_{_1}$ becomes nonzero and in analogy with the last section,
the conversion of kinetic to potential energy is achieved on a
time scale $\Lambda_{_2}=\xi_{_2}/\sqrt{K-1}$ as may be seen from:
\begin{align}
\xi_{_2}^2\ddot{\psi}_{_2}=\psi_{_2}\left[\eta
K-1+\psi_{_2}^2(1-K^2)\right]\,\text{ and }\,
\psi_{_1}^2=\eta-K\psi_{_2}^2.
\end{align}
 A remarkable feature is encountered when considering small
adsorption $\ell$; $V(\ell)$ is found not to exist when
$\ell<\ell_{_Q}$. A similar phenomenon was observed in the context of the interface potential for superconducting surface sheaths in Ginzburg-Landau theory~\cite{blossey}. This ``quantum" effect arises because, spatially
seen, the wave function $\psi_{_2}(z)$ is constrained to vanish at
$z=0$ and to have the value $\sqrt{1/K}$ at a certain point
$z=z_{_0}$; wave solutions therefore do not exist for too small
values of $z_{_0}$. Expression~\eqref{intpot2Wall} gives the
interface potential which is shown in Fig.~\ref{fig5} for $K=5$
and for different values of $\mu_{_2}/\overline{\mu}_{_2}$. The
quantum effect is apparent in Fig.~\ref{fig5} at the
thickness $\ell_{_Q}$, being the minimal thickness for which the
interface potential is defined.

Although we were unable to expand the potential in terms of large
$\ell$, we can point out its principal feature. For large film
thicknesses, the penetration depth of species $2$ into species $1$
(the bent path in Fig.~\ref{fig4}) saturates, such that the film
is grown from phase $2$, with $\psi_{_1}=0$ (the vertical path in
Fig.~\ref{fig4}). Therefore, the decay length of the
interface potential is $\xi_{_c}=\xi_{_2}/2$. Contrary to
findings of Sect.~\ref{Asmallcoherencelength}, $\xi_{_c}$ does not
change with $K$, and $|V(0)|$ increases for larger values
of $K$. Also, one can prove that for weak segregation, the
thickness $\ell_{_Q}$ (see Fig.~\ref{fig5}) diverges
logarithmically with $K-1$ such that the interface potential is
not well-defined when $K=1$.

\begin{figure}
\begin{center}
    \epsfig{figure=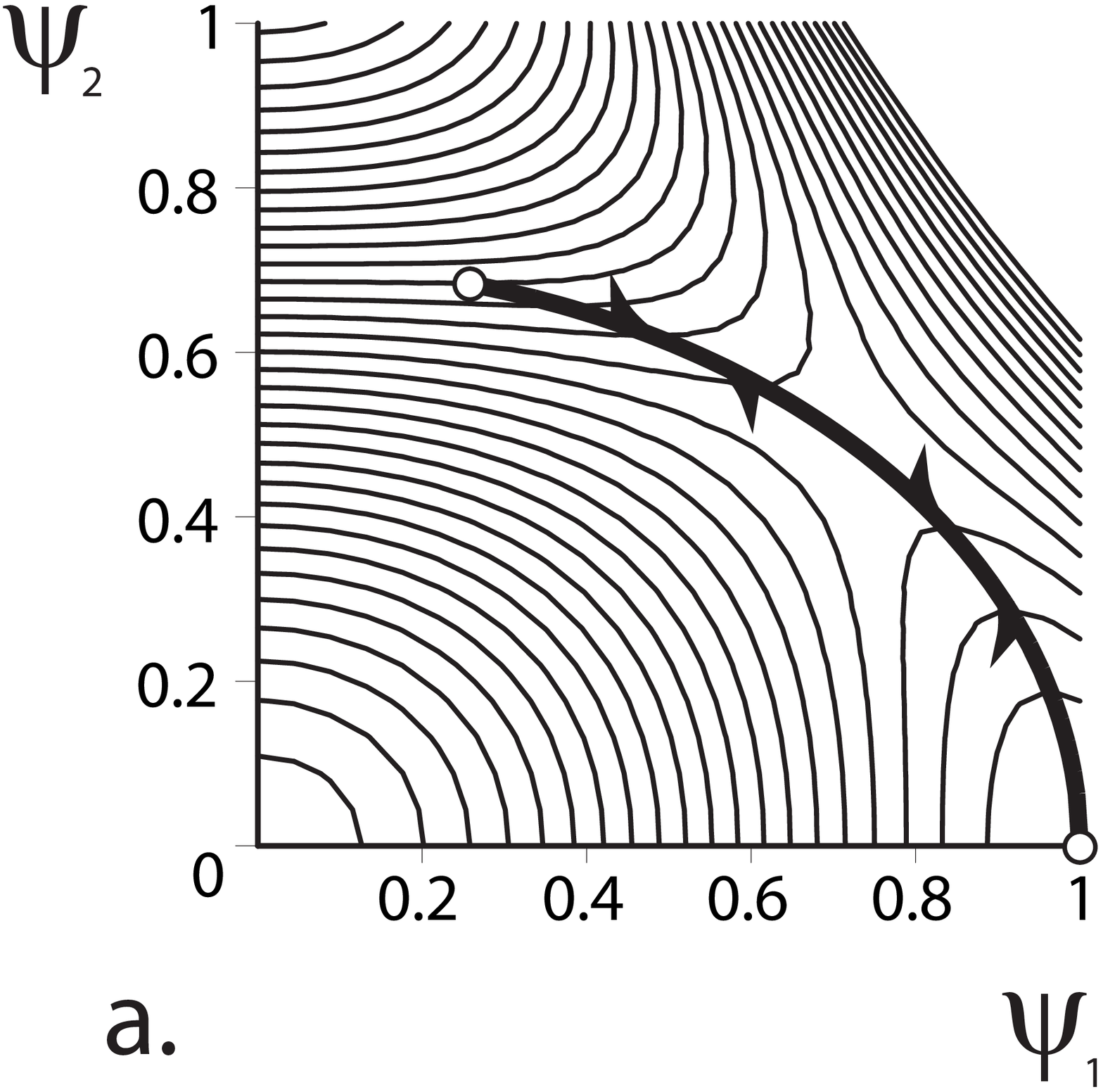,angle=0,width=120pt}
    \epsfig{figure=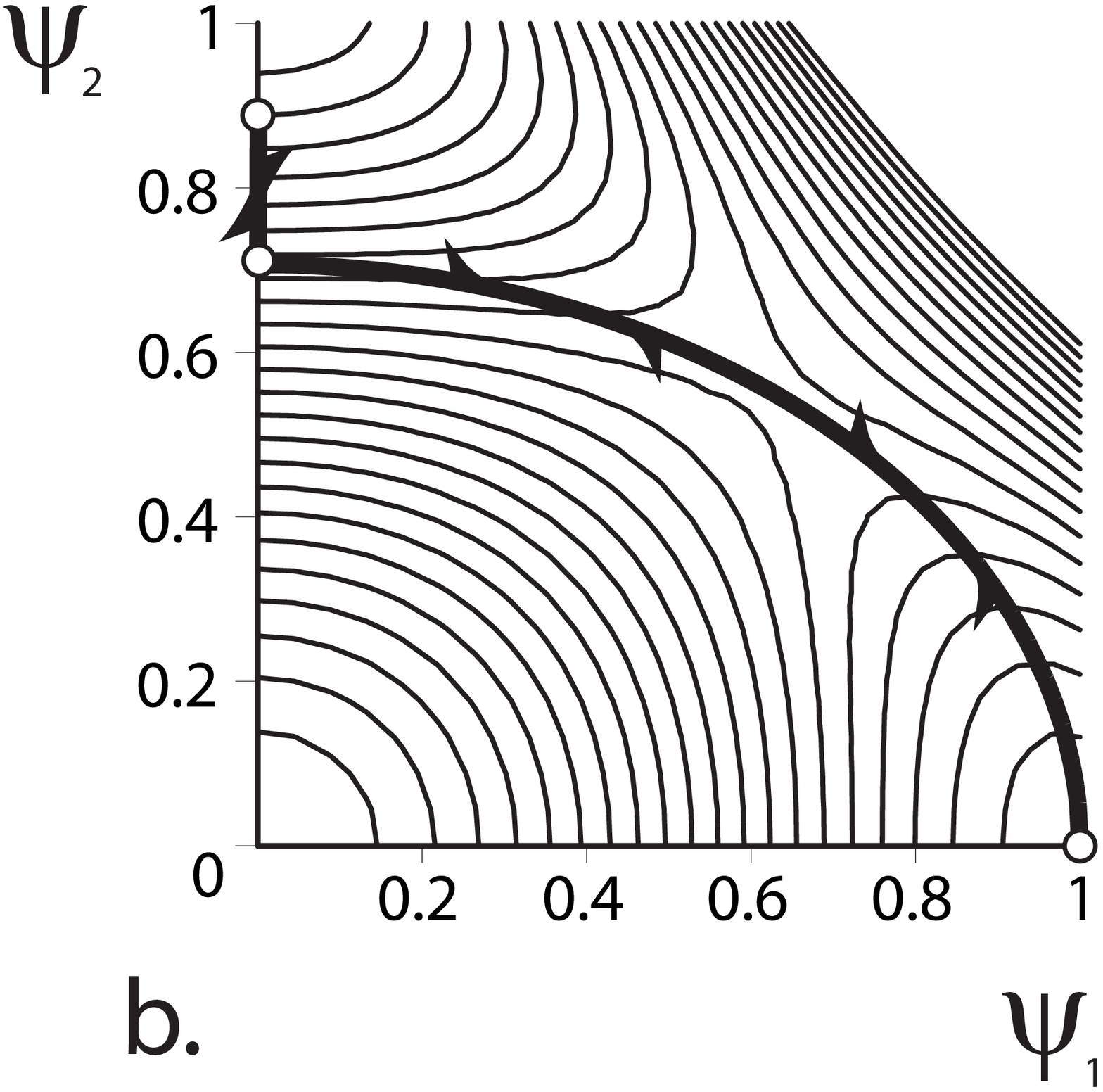,angle=0,width=120pt}
       \caption{The paths followed by the coordinates $(\psi_{_1},\psi_{_2})$
       in the case of ``bulk excitations'' when $\xi_{_1}/\xi_{_2}=0$, $K=2$ and
       $\eta=1.3$ (a) and $\eta=1.02$ (b). Both paths start in the point
       $(0,1)$ and finally return to this point.\label{fig6}
        }
\end{center}
\end{figure}

 The trajectories of the coordinates $(\psi_{_1},\psi_{_2})$ for ``bulk-nucleated states'' with a finite ``adsorption''
are shown in Fig.~\ref{fig6}a and~\ref{fig6}b. It is interesting
to note the existence of a stratification point $L$ for these
states (see Fig.~\ref{fig5}), where the intrusion of species $2$
is sufficient to sever space in two parts of pure phase $1$. In
Fig.~\ref{fig5}, we compare $V(\ell)$ with the excess energy
$\gamma$ of the bulk states which have the same ``adsorption'', by
subtraction of $\gamma_{_0}$. We found numerically that at
coexistence the bulk states have a lower energy for all
thicknesses $\ell$ and this is valid for all values of
$K$~\cite{footnote5}.

 When $\ell=\infty$, the condition for the excess energy of the
``bulk-nucleated states'' $\gamma_{_{1W}}+2\gamma_{_{12}}$ to be
lower than the excess energy of the ``wall-adsorbed states''
$\gamma_{_{12}}+\gamma_{_{2W}}$ yields the complete drying
condition $\gamma_{_{1W}}+\gamma_{_{12}}=\gamma_{_{2W}}$. In other
words, if pure phase $2$ were to be the bulk phase, pure phase $1$
would wet the wall. This was already encountered in
~\cite{indekeu}. Lastly, we note that it is not clear
whether or not the quantum effect disappears once we take
$\xi_{_1}/\xi_{_2}\neq 0$.

\begin{figure}
\begin{center}
    \epsfig{figure=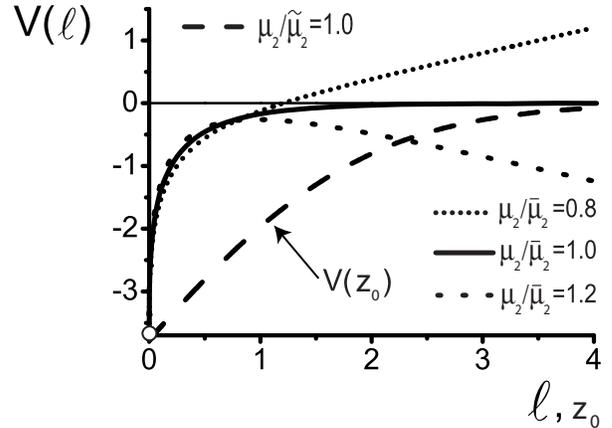,angle=0, height=170pt}
       \caption{Interface potential $V(\ell)$ in units of $P_{_1}\xi_{_2}$ as a function of the film thickness $\ell$ in units of
       $\xi_{_2}$ when $1/K=0$ and for different values of
       $\mu_{_2}/\overline{\mu}_{_2}$ as a function of both $\ell$ and $z_{_0}$ where the value of $z_{_0}$ is determined
       by the locus of the $1$-$2$ interface. The state of lowest energy corresponds
       to zero film thickness. Note the steep increase of the
       potential as a function of small $\ell$ whereas $V$
       varies linearly as a function of $z_{_0}$ for small values of $z_{_0}$. This behavior for small $\ell$ and small $z_0$ is due to the shift
       of the wave function profile of species $1$ with just a few particles of species $2$ being adsorbed at the surface.\label{fig7}
       }
\end{center}
\end{figure}
\subsection{Strong Segregation}\label{LargeInterPhaseRepulsionK}
The condition $1/K=0$ induces species $1$ and $2$ to have no
overlap such that the densities of both species vanish at a
certain distance $z_{_0}$ from the wall (at least whenever
$\ell\neq 0$). Since, in that case,
$\gamma_{_{12}}=\gamma_{_{2W}}+\gamma_{_{1W}}$, partial wetting
immediately follows from $\mathcal{S}<0$. Only density variations
of adsorbed phase $2$ modify the interface potential. In fact, 
``inflating" the wetting layer merely shifts the density profile
of species $1$, being
$\psi_{_1}=\tanh[(z-z_{_0})/(\sqrt{2}\,\xi_{_1})]$ through an increase
of $z_{_0}$. Species $2$ replaces species $1$ in the shifted
region, thereby modifying its own density profile. The exact interface
potential is provided in \eqref{intpot1} and can be seen in
Fig.~\ref{fig7} as a function of both $\ell$ and $z_{_0}$. For
large $\ell$, we derived that \cite{footnote6}:
\begin{align}\label{expans}
V(\ell)=h\ell
-32\sqrt{2}\,P_{_1}\xi_{_2}\,e^{-(4+\sqrt{2}\,\ell/\xi_{_2})}+\ldots
\end{align}
and $V(z_{_0})$ is found using the relation
$z_{_0}=\ell-2\sqrt{2}\,\xi_{_2}$, also valid for large $\ell$. Note that \eqref{expans} can
also be obtained by constraining the value of $z_{_0}$ instead of
$\ell$. In Fig.~\ref{fig7}, one observes an important difference
between $V(\ell)$ and $V(z_{_0})$ for small
values of $\ell$ and $z_{_0}$. This difference can be attributed
to a feature which is reminiscent of the quantum effect as
encountered in Sect.~\ref{Alargecoherencelength}. Indeed, wave
function $\psi_{_2}(z)$ is constrained to vanish at $z=0$ and at
$z=z_{_0}$. By application of a sufficiently large disjoining
pressure, solutions exist for all values of
$\mu_{_2}/\overline{\mu}_{_2}$. For a fixed small value of the
potential, the adsorption $\ell$ remains very small whereas the
corresponding value for $z_{_0}$ increases linearly. Therefore,
for low values of $z_{_0}$, the potential $V(z_{_0})$ roughly
quantifies the energy needed to push species $1$ to a distance
$z_{_0}$ from the wall and this is linear in $z_{_0}$.

\begin{figure}
\begin{center}
    \epsfig{figure=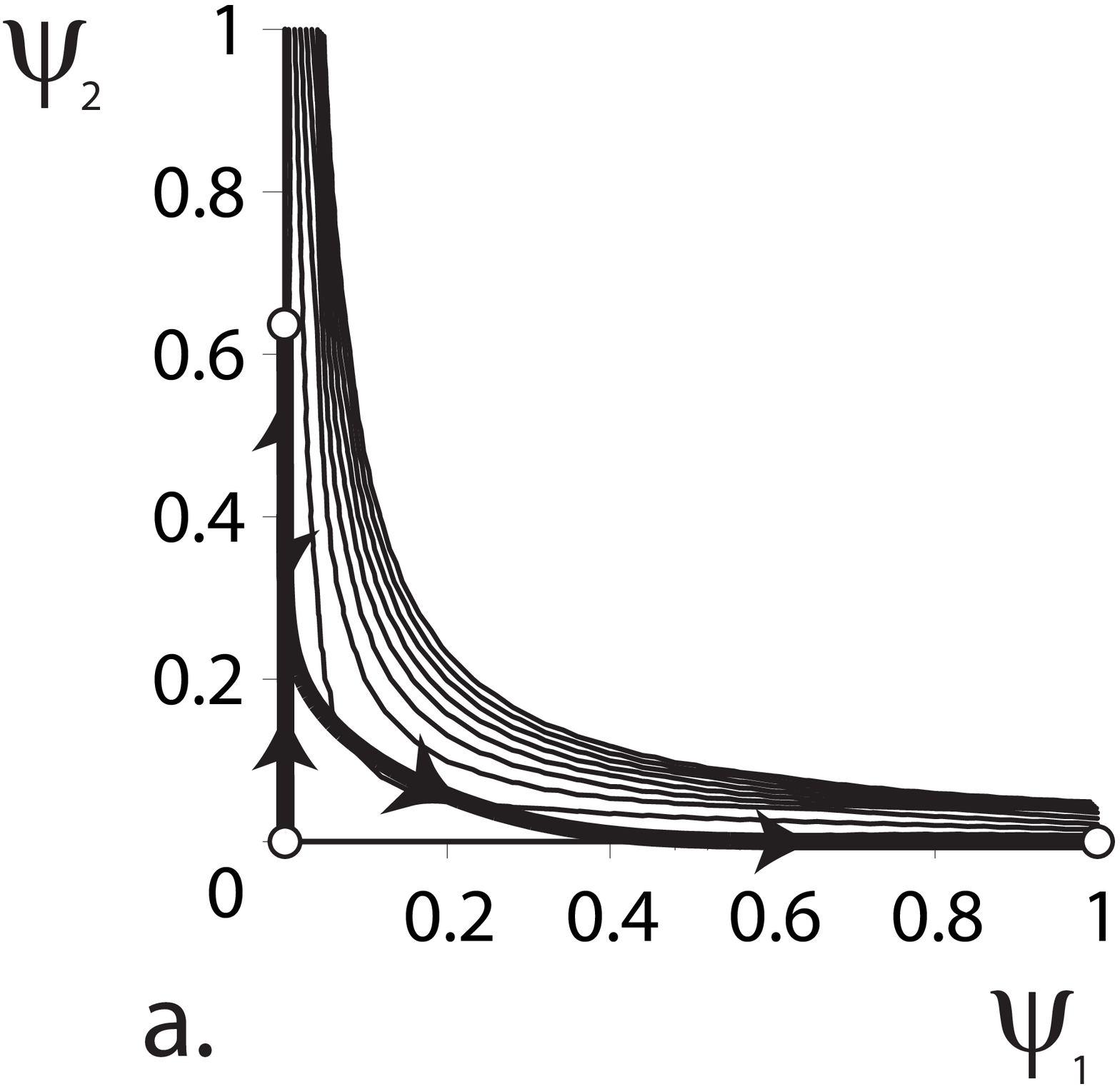,angle=0,width=115pt}
    \epsfig{figure=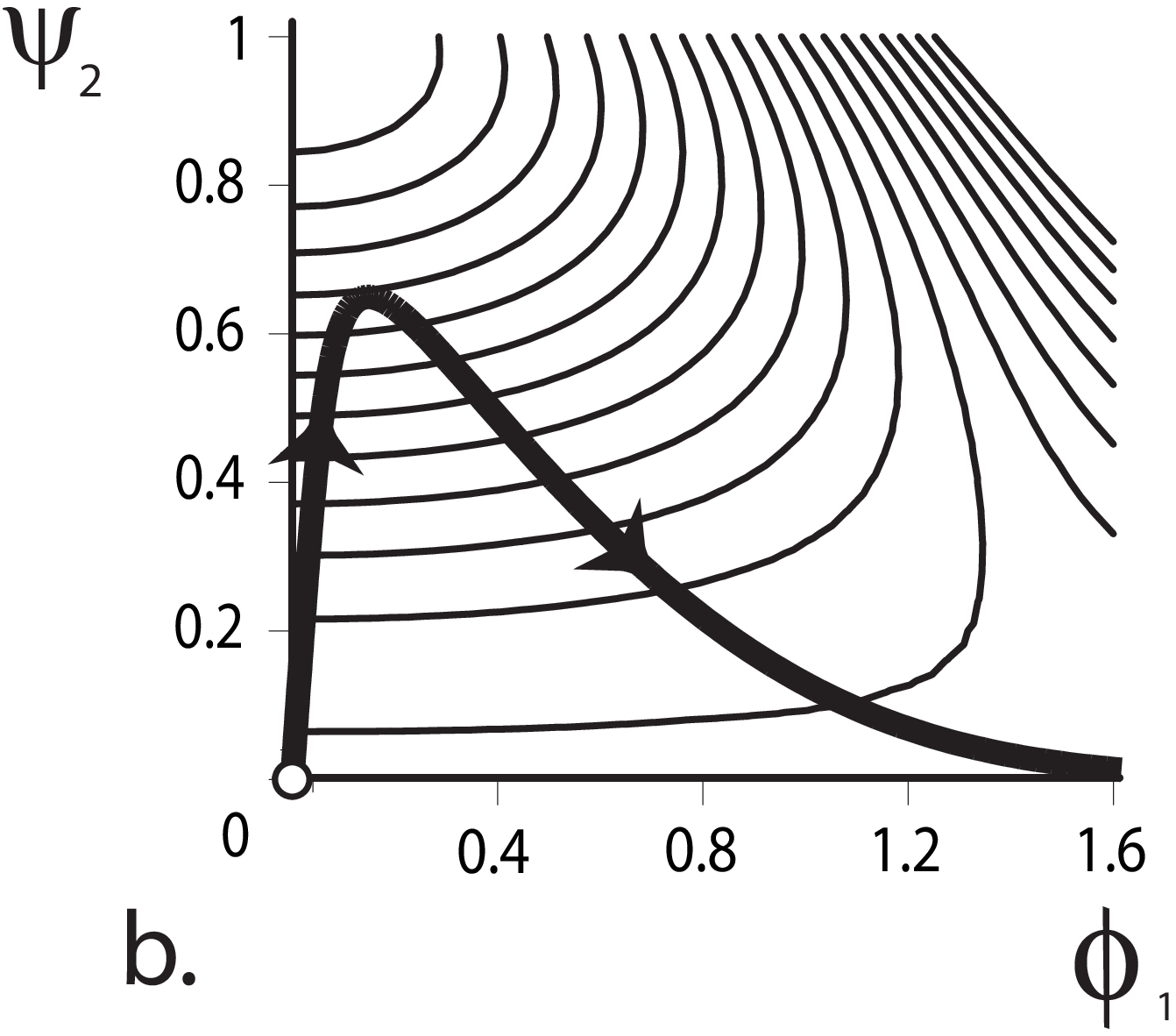,angle=0,width=125pt}
       \caption{a) The path for $(\psi_{_1},\psi_{_2})$ when $K=1000$ and
       $\eta=1.1$. In the upper right corner, we left out the
       equipotential curves due to their high density. b) A part of the path
       followed by $(\phi_{_1},\psi_{_2})$ for $\kappa=0.5$ and
       $\eta=1.4$. In the continuation of the depicted path,
       $\psi_{_2}\rightarrow 0$ while $\phi_{_1}$ increases.\label{fig8}
       }
\end{center}
\end{figure}

When $1/K$ is small but nonzero, an overlap region of the
condensates induces corrections to $\gamma_{_{12}}$ and therefore
$V(0)$ of order $1/\sqrt[4]{K}$. It is then possible to sketch the evolution of $\psi_{_1}$ and
$\psi_{_2}$ as was done for $K=1000$ in Fig.~\ref{fig8}a. One sees
the extreme deformation of the potential by dense contour lines
when both densities are nonzero.

In fact, the above analysis is only valid when
$1/\sqrt{K}\ll\xi_{_2}/\xi_{_1}$. In the following section we
study the intermediate region in which $\xi_{_2}/\xi_{_1}\approx
1/\sqrt{K}$ and small compared to unity.

\subsection{Both Strong Segregation and Strong Healing Length Asymmetry }\label{SmallxiandInfiniteK}
As indicated in Fig.~\ref{fig0}, a transition from PW to CW occurs
in regions E and D. These two regimes have in common the existence
of two important length scales; one length scale near the wall and
one length scale far from it, and therefore turn out more
difficult to solve, especially numerically. However, upon approach
of the points $(\xi_{_2}/\xi_{_1},1/K)=(0,0)$ and $(1,1)$, one
length will be much larger than the other such that the analysis
can be performed on two length scales separately.

 Next, we treat the case when both $\xi_{_2}\ll
\xi_{_1}$ and $1/K\ll 1$ while
\begin{equation}
\kappa\equiv \left[[\xi_{_2}/\xi_{_1}] \sqrt{K}\right] ^{-1},
\end{equation}
is of order unity~\cite{vanschaeybroeck}. In this limit and at
coexistence, PW is encountered when $\kappa < 3/\sqrt{2}$ while CW is found when $\kappa > 3/\sqrt{2}$; this may be deduced from \eqref{prewett}
and seen in Fig.~\ref{fig0}. The associated important length
scales are $\xi_{_2}$, near the wall and a much larger $\xi_{_1}$,
far from the wall. Whereas the largest energy contribution is
governed by density variations on the latter scale, the phenomena
on the former scale decide whether there is PW or CW.

Let us first focus on the density variations close to the wall. We
introduce the rescaling $\overline{z}\equiv z/\xi_{_2}$ and a new
wave function $\phi_{_1}$ as in \cite{vanschaeybroeck} following the calculational approach taken in \cite{boulter,boulter3}:
\begin{align}
\psi_{_1}=&\,[\xi_{_2}/\xi_{_1}]\left[\phi_{_1}-\frac{\Theta(\overline{z}-\delta)(\overline{z}-\delta)}{\sqrt{2}}\right]\\
&\quad+\Theta(z-\delta\xi_{_2})\tanh\left(\frac{z-\delta\xi_{_2}}{\sqrt{2}\xi_{_1}}\right),\nonumber
\end{align}
where $\phi_{_1}$ must have the asymptotic behavior
$\phi_{_1}(\overline{z}\rightarrow\infty)\sim (\overline{z}-\delta)/\sqrt{2}$
and $\phi_{_1}(\overline{z}\rightarrow-\infty)=0$. The scaling,
$\psi_{_1}\propto \xi_{_2}/\xi_{_1}$ stems from the fact that
$\psi_{_1}$ makes variations of order unity over a length $\xi_{_1}$
and hence variations of order $\xi_{_2}/\xi_{_1}$ over the length
scale $\xi_{_2}$. Close to the wall, one can expand $U$ and $T$ to
zeroth order in $\xi_{_2}/\xi_{_1}$ and in $1/K$:
\begin{figure}
\begin{center}
    \epsfig{figure=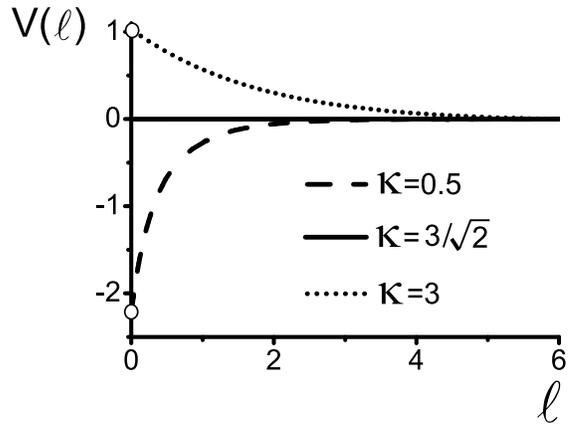,angle=0,width=210pt}
       \caption{
       The numerically obtained interface potential $V(\ell)$ in units of $P_{_1}\xi_{_2}$ as a function of the film thickness $\ell$ in units of
       $\xi_{_2}$ for the case $\xi_{_2}\ll \xi_{_1}$ and
       $1/K\ll 1$ while $\kappa=\left[[\xi_{_2}/\xi_{_1}]\sqrt{K}\right]^{-1}$ is finite. We see that
       one encounters CW for $\kappa >3/\sqrt{2}$ and PW for
       $\kappa < 3/\sqrt{2}$. These regimes are separated by the extraordinary flat
       potential. \label{fig9}
       }
\end{center}
\end{figure}
\begin{subequations}
\begin{align}
U[\phi_{_1},\psi_{_2}]&=2P_{_1}\left[\eta\psi_{_2}^{2}-
\frac{\psi_{_2}^{4}}{2}-\frac{\phi_{_1}^{2}
\psi_{_2}^{2}}{\kappa^2}\right]
,\\
T[\dot{\phi_{_1}},\dot{\psi_{_2}} ]&=2P_{_1}\left[
\dot{\phi_{_1}^2}+\dot{\psi^2_{_2}}\right],
\end{align}
\end{subequations}
 where the dot now indicates the derivative with
respect to $\overline{z}$. The extension to the region far from
the wall (where $z\gg\xi_{_2}$) is made by observing that there
$\psi_{_2}=0$ while $\phi_{_1}$ goes over into a $\tanh$ profile.
By straightforward calculation, one can write the interface
potential as:
\begin{align}\label{kappaint}
V(\ell)
=&\,4P_{_1}\xi_{_2}\int_{_0}^{_\infty}\text{d}\overline{z}\,\left[
\left(\dot{\phi}_{_1}-\frac{1}{\sqrt{2}}\right)^2+\dot{\psi}_
{_2}^{2}\right]\nonumber\\
&-\Pi \ell-2\delta P_{_1}\xi_{_2}-\gamma_{_{12}},
\end{align}
In Fig.~\ref{fig9} we depict the interface potential for different
values of $\kappa$. The results are obtained by a
numerical integration of the profiles, followed by the evaluation
of the potential with the functional~\eqref{kappaint}. Clearly we
see that the transition from PW to CW is \textit{mediated by a
completely flat potential for} $\kappa=3/\sqrt{2}$. The constancy of the interface potential is a property of crucial importance in the context of the wetting phase transition in this model. It corroborates the earlier observation of the infinite degeneracy of the grand potential at first-order wetting \cite{indekeu3}. This property is in stark contrast with the normally expected and ubiquitous double-minima structure of $V(\ell)$ at first-order wetting. 

We proceed by arguing that the interface potential in the limit
under consideration should have the form:
\begin{align}\label{haugs}
V(\ell)=h\ell
+Ce^{-\sqrt{2}\,\ell/\xi_{_2}}+De^{-2\sqrt{K}\,\ell/\xi_{_1}}+\ldots
\end{align}
where the amplitudes $C$ and $D$ are independent of $\ell$ and
positive such that no critical wetting transition is
possible~\cite{hauge}. First of all, for small values of $\kappa$
and thus $\xi_{_2}/\xi_{_1}\ll \sqrt{K}$, $V(\ell) - h\ell$, for large $\ell$,
should be the one obtained in Sect.~\ref{Alargecoherencelength}
and thus be proportional to $e^{-\sqrt{2}\,\ell/\xi_{_2}}$, which
means that $\xi_{_c}=\xi_{_2}/\sqrt{2}$. For large values of
$\kappa$, i.e., $\sqrt{K}\ll \xi_{_2}/\xi_{_1}$, we must obtain the
same result as obtained in Sect.~\ref{Asmallcoherencelength} which
is that $\xi_{_c}=\xi_{_1}/(2\sqrt{K})=\kappa\,\xi_{_2}/2$.
\begin{figure}
\begin{center}
     \epsfig{figure=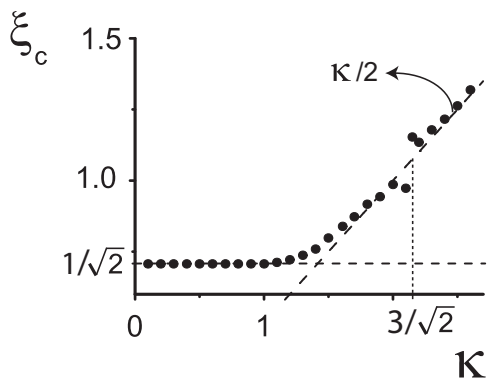,angle=0,width=180pt}
       \caption{
       Plot for the numerically obtained length $\xi_{_c}$ in units of $\xi_{_2}$ in the case
       of $\xi_{_2}\ll \xi_{_1}$ and $1/K\ll 1$. This length characterizes
       the exponential decay for large $\ell$. The transition from
        $\xi_{_c}=1/\sqrt{2}$ for low values of $\kappa$ to
        $\xi_{_c}=\kappa/2$ is obvious. For values of $\kappa$ lower than
        $3/\sqrt{2}$, we are in the PW regime whereas for higher values,
        CW occurs.
        \label{fig11}
       }
\end{center}
\end{figure}
These considerations are supported by fitting the numerically
obtained values for $\xi_{_c}$ as a function of $\kappa$, as shown
in Fig.~\ref{fig11}. One observes that, indeed, $\xi_{_c}$ takes
the value $\xi_{_2}/\sqrt{2}$ for low $\kappa$ and
the value $\kappa\,\xi_{_2}/2$ for larger values of $\kappa$ \cite{footnote7}.  Note that for different types of surfaces (involving softer walls) also critical wetting transitions have been established in GP theory \cite{vanschaeybroeck2}.

\subsection{Weak Segregation}\label{weaksegregationsection}
We now direct our attention to the case when both $0< K-1\ll 1$ and
$0< \xi_{_1}/\xi_{_2}-1\ll 1$. In this regime we will show that the transition from PW
to CW at coexistence, mediated by a flat interface potential,
takes place when $\sqrt{K-1}=2\sqrt{2}(\xi_{_1}/\xi_{_2}-1)/3$, as is readily
deduced from \eqref{prewett}. In the following, we take the
Lagrange multiplier to be
$\Pi=2P_{_1}\mu_{_2}/\overline{\mu}_{_2}$; in order to fix the
film thickness, we use instead a parameter which is featured in
our expansion. The expansion parameters are $\sqrt{K-1}\ll 1$ and
$\xi_{_1}/\xi_{_2}-1\ll 1$ and we assume both to be of the same
order. We extend a method, used earlier by Malomed et
al.~\cite{malomed} and Mazets~\cite{mazets} for the calculation of
the interfacial tension. In the limit $K\rightarrow 1$, the
potential $U$ of \eqref{energies} is rotationally invariant;
it is then natural to rewrite $\psi_{_1}$ and $\psi_{_2}$ as
follows\label{chidef}:
\begin{subequations}
\begin{align}
\psi_{_1}\equiv\,g(z)\sin[\chi(z)],\\
\psi_{_2}\equiv\,g(z)\cos[\chi(z)],
\end{align}
\end{subequations}
where the boundary conditions~\eqref{bound} imply
$\left.g(z)\right|_{z=0}=0$, $\left.\chi(z)\right|_{z=0}<\pi/2$,
$\left.\chi(z)\right|_{z=\infty}=\pi/2$ and
$\left.g(z)\right|_{z=\infty}=1$. The existence of two largely
different length scales $\xi_{_2}$ and $\xi_{_2}/\sqrt{K-1}$ gives
rise to an expansion in its most general form:
\begin{figure}
\begin{center}
    \epsfig{figure=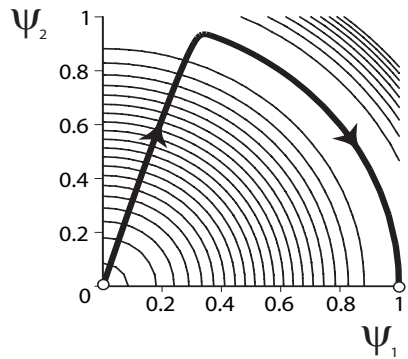,angle=0,width=150pt,height=140pt}
       \caption{The path for the
        normalized densities in the case of weak segregation with
       $K=\xi_{_1}/\xi_{_2}=1.001$ and $\chi(0)=0.3$. Notice the
       (near-)rotational symmetry of the underlying potential $U$.\label{fig12}
       }
\end{center}
\end{figure}
\begin{subequations}
\begin{align}
\chi(z)=&\chi_{_0}(z)+\chi_{_1}(z\sqrt{K-1}
)\\
&+\sqrt{K-1}\,\chi_{_2}(z)\ldots,\nonumber\\
g(z)=&g_{_0}(z)+g_{_1}(z\sqrt{K-1} )+\sqrt{K-1}\,
g_{_2}(z)\\
&\quad+\sqrt{K-1}\, g_{_3}(z\sqrt{K-1})+\ldots\nonumber
\end{align}
\end{subequations}
Here, $\chi_{_1}$, $g_{_1}$ and $g_{_3}$ vary as a function of the
``slow coordinate'' $z\sqrt{K-1} $ whereas $g_{_0}$, $g_{_2}$,
$\chi_{_2}$ and $\chi_{_0}$ vary on the short length scale
$\xi_{_2}$ near the wall. Expanding the modified GP equations in the small parameters $\sqrt{K-1}$ and $\xi_{_1}/\xi_{_2}-1$ and equating the same
orders of magnitude, leads to:
\begin{subequations}
\begin{align}
\chi_{_0}=g_{_1}=&g_{_3}=0,\\
\xi_{_2}^2\ddot{g}_{_0}=-&g_{_0}+g_{_0}^3,\\
\xi_{_2}^2\ddot{g}_{_2}=g_{_2}[-1+3g_{_0}^2]&-2\left(\frac{\xi_{_1}/\xi_{_2}-1}{\sqrt{K-1}}\right)\xi_{_2}^2\ddot{g}_{_0}\sin[\chi_{_1}(0)],\\
2\dot{g}_{_0}\dot{\chi}+\ddot{\chi}_{_2}=-&(\xi_{_1}/\xi_{_2}-1)\ddot{g}_{_0}\sin^2[2\chi_{_1}(0)],\\
\xi_{_2}\dot{\chi}_{_1}=-&\sqrt{K-1}\,\frac{\sin 2\chi_{_1}}{2}.
\end{align}
\end{subequations}
The last expression is obtained by setting $g=1$ and taking the
first-order terms of the subtraction of the GP equations. In the third
and fourth expression, we expand
$\chi_{_1}(z)=\chi_{_1}(0)+\sqrt{K-1}\,\chi_{_2}(z)+\mathcal{O}(K
-1)$ about its value at the wall; this is justified since
$g_{_2}$ is nonzero only near the wall. The solutions are (with
$K\neq 1$):
\begin{figure}
\begin{center}
    \epsfig{figure=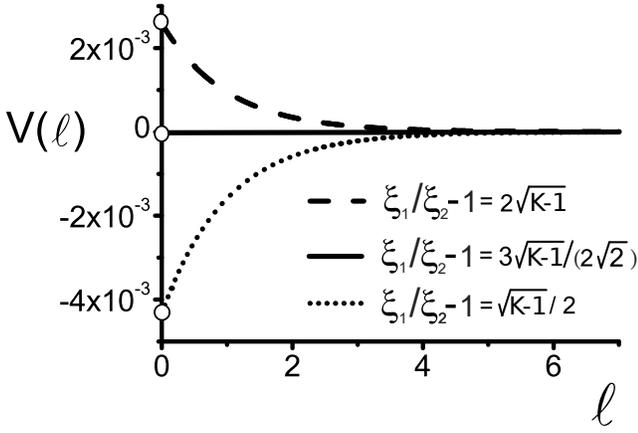,angle=0, height=160pt}
       \caption{Interface potential $V(\ell)$ in units of $P_{_1}\xi_{_2}$ as a function of the film thickness $\ell$ in units of
       $\xi_{_2}/(2\sqrt{K-1})$ for weak segregation and small $\xi_{_1}/\xi_{_2}-1$.
       Here, $\sqrt{K-1}=0.01$ while $\xi_{_1}/\xi_{_2}-1$
       varies. At the point where $\xi_{_1}/\xi_{_2}-1=3\sqrt{K-1}/(2\sqrt{2})$, we go over from
       PW to CW through a flat potential. \label{fig13}
       }
\end{center}
\end{figure}
\begin{subequations}
\begin{align}
g_{_0}(z)&=\tanh[z/(\sqrt{2}\,\xi_{_2})],\\
g_{_2}(z)&=\left(\frac{\xi_{_1}/\xi_{_2}-1}{\sqrt{K-1}}\right)\sin^2[\chi_{_1}(0)](g_{_0}^2-1)\text{arctanh}\,g_{_0},\\
\chi_{_1}(z)&=\text{arctanh}\left(e^{-\sqrt{K-1}\,
z/\xi_{_2}}\tanh[\chi_{_1}(0)]\right).
\end{align}
\end{subequations}
The solution for $\chi_{_2}$ is irrelevant for our further
analysis. The value of $\chi_{_1}(0)$ is undetermined and we use
this parameter to tune the film thickness. The interface potential
can now be separated into a part close and a part far from the
wall. It is interesting that, independently of the film
thickness, both density profiles vary on the small length
scale $\xi_{_2}$ close to the wall, whereas a length scale
$\xi_{_2}/\sqrt{K-1}$ is found for the behavior far from the
wall. Note that this set of equations is consistent with an
expansion of the conservation of energy.

By use of \eqref{consenergy} and \eqref{simpleform2}, we then
find to first order in $\sqrt{K-1}$:
\begin{align}
\frac{V(\ell)+\gamma_{_0}-h\ell}{2P_{_1}}=&\int_{_0}^{_\infty}\text{d}z\left[(1-g_{_0}^2)^2+(K
- 1)\frac{\sin 2\chi_{_1}}{4}\right.\nonumber\\
&\left.+4\sqrt{K-1}\, g_{_0}g_{_2}(g_{_0}^2-1)\right]+\ldots,
\end{align}
while the expression for the film thickness is
\begin{align}
\ell=\int_{_0}^{_\infty}\text{d}z\left[
(g_{_0}^2-1)\cos^2[\chi_{_1}(0)]+\cos^2\chi_{_1}\right]+\mathcal{O}(\sqrt{
K - 1 }).
\end{align}
Both $\ell$ and $V(\ell)$ depend on $\chi_{_1}(0)$ in such a way that this
parameter can be eliminated. One can derive the exact interface
potential to first order in $\sqrt{K-1}$ and to first order in $\xi_{_1}/\xi_{_2}-1$, which we give in
\eqref{intpot4}; for large $\ell$ it reduces to:
\begin{align}\label{weaksegpot}
V(\ell)=&h\ell+2P_{_1}\xi_{_2}
\left[\frac{2\sqrt{2}}{3}(\xi_{_1}/\xi_{_2}-1)-\sqrt{K-1}\right]\nonumber\\
&\times\exp\left(-\frac{ 2\ell\sqrt{K-1}}{\xi_{_2}}\right)+\ldots
\end{align}
Obviously, when $h=0$ and
\begin{align}\label{conditie}
\sqrt{K-1}=2\sqrt{2}(\xi_{_1}/\xi_{_2}-1)/3,
\end{align}
this interface potential, as well as the exact one in \eqref{intpot4}, vanishes for all $\ell$. Note that here,
as opposed to the case in Sect.~\ref{SmallxiandInfiniteK}, only
one length scale, namely $\xi_{_2}/(2\sqrt{K-1})$, determines the
interface potential. The derived interface potential is depicted
in Fig.~\ref{fig13} for different values of $\xi_{_1}/\xi_{_2}-1$.
Clearly, one can have either PW or CW and there is a completely
flat interface potential when \eqref{conditie} is satisfied.

\section{Line Tension}\label{linetension}
A system having a three-phase contact line can be attributed an
excess energy which is proportional to the contact length. The excess energy per
unit length is called the \textit{line tension}. It is important to appreciate that a 
line tension can be of either sign, it need not be positive. The prerequisite for such contact
line to be present is a finite (non-zero) contact angle, that is, to have a
partial wetting state, or, to be just at a first-order wetting
point. In this work, the latter was encountered in the crossover
regions of Sect.~\ref{SmallxiandInfiniteK} and
Sect.~\ref{weaksegregationsection}. In the case of
an {\em ordinary first-order wetting} transition the line tension at the first-order wetting transition can be approached
along two paths: along the coexistence line (from PW) and along the
first-order prewetting line (PL) off of bulk coexistence. Along the latter path the line tension is referred to as boundary tension, because the adsorbate is a single phase and there is consequently no three-phase contact line. The boundary tension at a
first-order thin-thick transition (along an ordinary PL) can be seen as the
energy cost per unit length of the density inhomogeneity formed
between a thin and a thick film. The boundary tension is in fact an interfacial tension in an effectively $d-1$-dimensional subset of a system with bulk dimensionality $d$ and is therefore non-negative. Along a thin-thick transition
line, the boundary tension starts off from zero at the prewetting critical point
and typically (for short-range forces) increases to a finite positive value at the wetting transition at coexistence. For long-range forces also a divergence of the line (and boundary) tension at wetting is possible~\cite{indekeu4}. 

However, in our system the wetting phenomena are extraordinary and do not follow the typical behavior and this has surprising consequences also for the line tension. In our case, the prewetting line PL is entirely critical and the first-order character only shows up at one point, which is the wetting transition at coexistence. This implies that the boundary tension is zero along PL, because there is no density jump across PL at all. Moreover, also at the wetting point where PL meets two-phase coexistence, the line tension is zero. This property of a vanishing line tension at wetting 
follows from the fact that the interface potential is perfectly flat at the wetting transition (there is no barrier in $V(\ell)$). This is consistent with the infinite degeneracy of the grand potential at wetting first noticed in \cite{indekeu3}.

\begin{figure}
\begin{center}
    \epsfig{figure=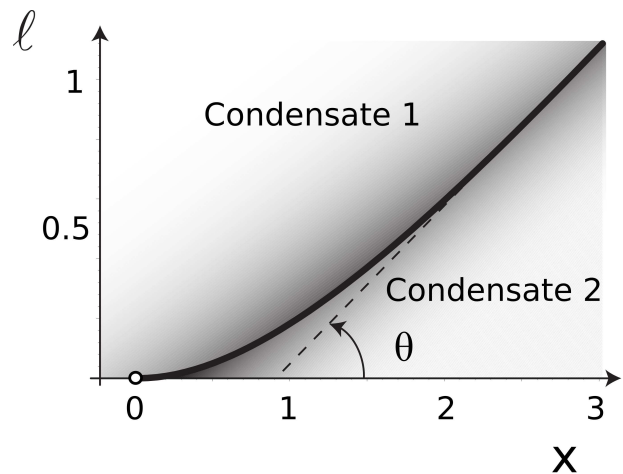,angle=0, width=230pt}
       \caption{Cross-section perpendicular to the three-phase contact line where phases $1$, $2$ and the wall meet. Depicted is the interface displacement $\ell(x)$ in units of $10\,\xi_{_c}$ against the coordinate position $x$ in units of $\xi_{_c}$
        for a $1$-$2$ interface which is incident on a hard
        wall and for an interface potential which is of the form
        $V(\ell)=\mathcal{S}\,e^{- 0.1\, \ell/\xi_{_c}}$. For this calculation the ratio $-\mathcal{S}/\gamma_{12}$ equals 0.2. The interface meets the wall at $x=0$ (open circle) tangentially with zero slope and $\ell(x) \propto x^2$ for small $x>0$ (while $\ell (x) =0$ for $x<0$).
       \label{fig14}
       }
\end{center}
\end{figure}
We focus on a three-phase contact line the cross-section of which is shown in Fig.~\ref{fig14}. The line is centered about $x=0$ and $\ell =0$ ($\ell =0$ coincides with the surface of the optical hard wall) and runs along the $y$-direction (perpendicular to the figure). The interface displacement $\ell(x)$ has translational symmetry in the $y$-direction. Note that the shape of $\ell(x)$ displays a monotically increasing slope. This feature would be typical for a transition zone that corresponds to the approach to an ordinary critical wetting transition rather than a first-order wetting transition. Upon approach of an ordinary first-order wetting transition a transition zone with an  inflection point would be expected \cite{indekeu1992}. This illustrates once again the extraordinary character of the wetting transition in this model of adsorbed BEC mixtures. 

According to the interface displacement model (IDM) \cite{indekeu1992,dobbs}, the line tension
$\tau$ is the following functional of the interface displacement $\ell(x)$, at bulk two-phase coexistence,
\begin{align}\label{linetensdef1}
\tau[\ell]=&\int_{_{-\infty}}^{_\infty}\text{d}x\,\left[\gamma_{_{12}}\left[\sqrt{1+\ell_{_x}^2(x)}-1\right]\right.\nonumber\\
&+V(\ell(x))-\mathcal{S}+c(x)\Big].
\end{align}
where $\ell_{_x}=\text{d}\ell(x)/\text{d} x$ and the piecewise
constant $c(x)$ is such that the integrand vanishes at large
values of $x$. Note that the $V(\ell)$ used in \cite{indekeu1992,dobbs} is shifted with respect to our $V(\ell)$ by a constant equal to the spreading coefficient $S$. The first term in \eqref{linetensdef1} measures the excess energy per unit
length due to the surface curvature close to the three-phase
contact line. By a minimization procedure, one derives the relevant Euler-Lagrange equation and associated constant of the motion. One finds that the equilibrium (or optimal) $\ell (x)$ starts from zero, at, say $x=0$, with zero slope and finite second derivative, as shown in Fig.~\ref{fig14}. Note that $\ell (x) =0$ for all $x<0$ (there is no microscopic adsorbed film of condensate 2 in this region). The line tension functional evaluated in this optimal profile provides the equilibrium line tension and can be written as the following integral~\cite{dobbs}:
\begin{align}\label{linetensdef2}
\tau=\sqrt{2\gamma_{_{12}}}\int_{_0}^{_\infty}\text{d}\ell\,&\left[
\sqrt{(V(\ell)-\mathcal{S})\left(1-\frac{V(\ell)-\mathcal{S}}{2\gamma_{_{12}}}\right)}\right.\nonumber\\
&\quad\left.-
\sqrt{-\mathcal{S}\left(1+\frac{\mathcal{S}}{2\gamma_{_{12}}}\right)}\right]
\end{align}

The validity of the use of the IDM for the purpose of calculating the properties of the three-phase contact line is determined by the requirement that $\ell (x)$ be slowly varying with $x$. Indeed, the $V(\ell)$ that is employed is a quantity that is {\em a priori} calculated for a uniform thickness $\ell$, not a spatially varying one. Nevertheless, extending it to a spatially varying function $V(\ell (x))$, one may hope to get a reasonable approximation for inhomogeneous configurations such as those we are interested in here. However, for contact angles that are not small, and certainly for $\theta \approx \pi/2$, the gradient $d \ell/dx$ is too large (and even diverges) for this model to be reliable and unphysical features should be expected. A discussion of some of the artifacts that may result can be found in \cite{dobbs}.

The interface potentials as we found here in the GP theory, were all, except for
one, of the typical form, asymptotically for $\ell \rightarrow \infty$,
\begin{equation}
\label{Vtypical}
    V_{\rm typ}(\ell)=\mathcal{S}\,e^{-\ell/\xi_{_c}},
\end{equation} for partial wetting states ($\mathcal{S} \leq 0$) at
two-phase coexistence. We will take this $V_{\rm typ}(\ell)$ as our model interface potential in what follows and attempt to determine the corresponding line tension and interface displacement as accurately as possible, within the IDM and also beyond the IDM by means of a conjecture based on symmetry and analyticity considerations. 

Within the IDM, substitution of \eqref{Vtypical} in \eqref{linetensdef2} brings us to
the analytic expression for the line tension:
\begin{align}\label{linetensdef3}
\tau=\frac{\gamma_{_{12}}\xi_{_c}}{2}&\left[(2\varsigma-1)(2\text{arcsin}[2\varsigma-1]+\pi)\right.\nonumber\\
&\quad\left.+4\sqrt{\varsigma(1-\varsigma)}(\ln[4(1-\varsigma)]-1)\right].
\end{align}
where
$2\varsigma=-\mathcal{S}/\gamma_{_{12}}=1-\cos\theta$\label{defvarsigma} 
with $\theta$ the angle at which the asymptote (for $x\rightarrow \infty$) to the wedge is incident on the
wall (see the dashed line in Fig.~\ref{fig14}). 

\begin{figure}
\begin{center}
    \epsfig{figure=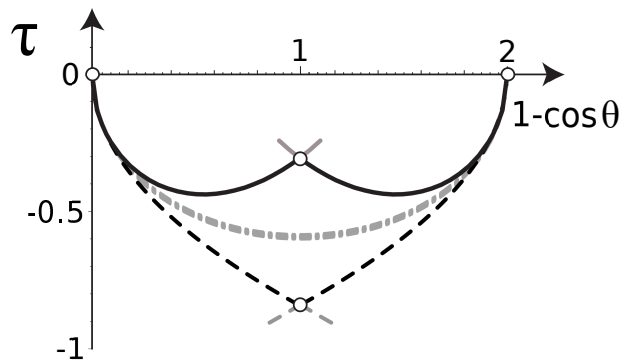,angle=0, width=230pt}
       \caption{
        The line tension $\tau$ in units of $\gamma_{_{12}}\xi_{_c}$
        versus $1-\cos\theta$, where $\theta$ is the interface
        inclination angle defined asymptotically, far from the three-phase contact line (i.e., the contact angle as predicted by
        Young's law). The lower curves (dashed lines) correspond to the gradient-squared approximation within the IDM.  The upper curves (solid lines) correspond to the full IDM calculations. For both approaches, the physically stable part is in black (the extension in gray). The smooth curve (dash-dotted in gray), which runs in between the stable parts of the IDM curves, corresponds to the conjecture \eqref{conjecture} for the exact line tension in GP theory consistent with the typical interface potential \eqref{Vtypical}.
       \label{fig15}
       }
\end{center}
\end{figure}

This expression for $\tau$ is displayed in Fig. \ref{fig15} as the solid line. Note that the model is ``mirror" symmetric about $\theta = \pi/2$ (wetting/drying symmetry) and all properties at $\theta$ are identical to those at $\pi-\theta$ provided the roles of phases (and species) 1 and 2 are interchanged. This explains the presence of two solid lines, each of which is the supplement of the other. Since the IDM  is most reliable for weakly varying $\ell(x)$, the reliable parts are those in black and the extensions that do not correspond to the physically stable solutions are in gray. The singularity at the crossing point $\theta = \pi/2$ is an artifact of the IDM. There is no physical singularity at the ``neutral" point at which there is no preferential adsorption of one of the phases. On the other hand, the singularities at $\theta = 0 $ (and $\pi$) are physical and correspond to the wetting (and drying) phase transitions. Before we turn to those in more detail we point out yet another interesting fact.

An often used simplification of the IDM consist of expanding the square root in \eqref{linetensdef2} to first order in the gradient squared, thereby explicitly acknowledging that the model is meant to serve (only) for weakly varying $\ell (x)$. This is the so-named gradient-squared approximation of the IDM. At the level of \eqref{linetensdef2} one easily verifies that the gradient-squared approximation amounts to reducing the integrand in \eqref{linetensdef2} to $\sqrt{V(\ell)-\mathcal{S}} -
\sqrt{-\mathcal{S}}$. In this approximation the analytic result for $\tau$ is the following simple expression
\begin{equation}
\label{gradient squared tau}
\tau = -2 \sqrt{2}\,(1-\ln 2)\,
\gamma_{_{12}}\xi_{_c}\,\sqrt {1 - \cos\theta},
\end{equation}
This simplification of \eqref{linetensdef3}, together with its symmetric supplement, is displayed as the dashed lines in Fig.~\ref{fig15}. While \eqref{gradient squared tau} is expected to lose accuracy more rapidly than \eqref{linetensdef3} upon increasing $\theta$ from zero, both are expected to be equally precise for small $\theta$ and indeed the asymptotic forms of \eqref{gradient squared tau} and \eqref{linetensdef3} approaching wetting are coincident. Specifically, $\tau$ approaches zero with a square-root singularity in the variable $1 - \cos \theta$ (see Fig.~\ref{fig15}). Consequently, approaching complete wetting, for $\theta\rightarrow0$, we find that the
line tension is asymptotically equal to 
\begin{equation}
\label{linearized}
\tau \sim -2 \,(1-\ln 2)\,
\gamma_{_{12}}\xi_{_c}\,\theta,
\end{equation}
and thus $\tau$ at wetting approaches zero from negative values.

This result is surprising because it is reminiscent of the behavior of the line tension close to critical wetting in systems with short-range forces (i.e., exponentially decaying $V(\ell)$) in a standard mean-field theory \cite{indekeu1992}. Since we are dealing with a wetting transition that is not critical but of first order, $\tau$ is expected to attain a non-zero and finite positive value, also from below, at wetting \cite{indekeu1992}. The behavior of the line tension at wetting depends, in mean-field theory, mainly on two characteristics. One is the order of the transition and the other is the range of the forces. Beyond mean-field theory there are fluctuation effects. For a review see \cite{indekeu4}. 

We conclude that the line tension for adsorbed BEC mixtures displays a hybrid character, which is caused by the extraordinary absence of a barrier in the interface potential $V(\ell)$. The absence of a barrier implies that  for
$\theta\rightarrow 0$ there is no transition zone which builds up
in $\ell(x)$ between zero thickness and a macroscopic (infinite) thickness. Instead, a completely flat profile $\ell(x)$
results. Note that we cannot exclude that physically a transition zone at first-order wetting may still exist in this system, but an interface displacement model based on $V(\ell)$ cannot capture it.

The results for $\tau$ obtained within the IDM, in gradient-squared approximation, \eqref{gradient squared tau},  and beyond this approximation,  \eqref{linetensdef3}, suggest a conjecture for the exact solution for  $\tau$ within GP theory and for the simple choice of interface potential given by \eqref{Vtypical}, without correction terms that become important at small $\ell$ and would modify also the line tension results quantitatively. This conjecture is based on three assumptions: i) there is nothing physically special about $\theta = \pi/2$ (obtained for $\xi_{_1} = \xi_{_2}$) and $\tau$ must be smooth (i.e., analytic) in that vicinity, ii) the $\theta \rightleftharpoons \pi - \theta$ symmetry must be respected, and iii) the asymptotic behavior near wetting (and drying) established with the help of the IDM calculations must be preserved in detail. The simplest function which satisfies i)-iii) is:
\begin{equation}
\label{conjecture}
\tau_{\rm conj} = -2 \,(1-\ln 2)\,
\gamma_{_{12}}\xi_{_c}\,\sin\theta,
\end{equation}
and it is displayed by the dash-dotted line in gray in Fig. \ref{fig15}. Perhaps it is possible to verify this conjecture (to a decisive extent) by designing an exact calculation of $\tau$ right at $\theta = \pi/2$ in GP theory. 

Let us now return to the IDM results and close this section with some remarks. While the expression \eqref{linetensdef3} is general and displays that $\tau$ is essentially a numerical factor times the interfacial tension multiplied by a characteristic surface-related length, it is possible in special cases (cf. the different regimes we studied) to obtain an explicit dependence on other characteristic lengths and on the interaction strength in our system. For example, in Sect.~\ref{SmallxiandInfiniteK}, we considered the case
$\xi_{_2}/\xi_{_1}\rightarrow 0$ and $K\rightarrow \infty$ while
$[\xi_{_2}/\xi_{_1}] \sqrt{K}$ was of order unity. We analyzed
numerically the length scale of exponential decay of the interface
potential, see \eqref{haugs}, and the result was shown in
Fig.~\ref{fig11}. For the partial wetting regime, i.e.,~when
$[\xi_{_2}/\xi_{_1}] \sqrt{K}>\sqrt{2}/3$, we found that
$\xi_{_c}$ goes over from the value $\xi_{_2}/2$ to
$\xi_{_1}/(2\sqrt{K})$ upon varying $[\xi_{_2}/\xi_{_1}] \sqrt{K}$. The dependence of the line tension on $\xi_{_1}$, $\xi_{_2}$ and $K$ therefore obeys:
\begin{align}\label{linetens2}
\tau\propto  P_{_1}\xi_{_1}\xi_{_c},
\end{align}
with a proportionality factor of order unity, and 
where we used that $\gamma_{_{12}}\approx P_{_1}\xi_{_1}$ and
where the value of $\xi_{_c}$ is plotted in Fig.\ref{fig11}. Note
that $\tau$ scales as $1/\sqrt{K}$ and is therefore small.

We consider now the case of weak segregation (see
Sect.~\ref{weaksegregationsection}), i.e.,~when
 $0< \xi_{_1}/\xi_{_2}-1\ll 1$ and $0< K-1\ll 1$, for which \eqref{weaksegpot} expresses the exact interface potential, to leading order for large $\ell$.  At
PW or when   $2\sqrt{2}[\xi_{_1}/\xi_{_2}-1]/3<\sqrt{K-1}$, we found
that 
\begin{align}
\cos\theta=\frac{2\sqrt{2}}{3}\frac{[\xi_{_1}/\xi_{_2}-1]}{\sqrt{K-1}},\quad
\xi_{_c}=\frac{\xi_{_2}}{2\sqrt{K-1}},
\end{align}
since $\gamma_{_{12}}= 2P_{_1}\xi_{_{2}}\sqrt{K-1}$. 

\section{Conclusion}\label{interface potentialconclusion}
We established the interface potential $V(\ell)$ for binary
mixtures of Bose-Einstein condensates near a hard wall. The
interface potential relates a configuration of adsorbed film
thickness $\ell$ of species $2$ to its excess grand potential per
unit area such that the equilibrium thickness is the value which
minimizes the potential. Generally, we find for large $\ell$, $V(\ell)=h\ell+Ae^{-\ell/\xi_{_c}}+\ldots$
where $h$ is the bulk field. At two-phase coexistence ($h=0$), the leading exponential decay dominates the entire $V(\ell)$. There is no barrier (contrary to what happens in other models where the next-to-leading terms may be relevant, too). When $A$ is positive, complete wetting (CW) occurs whereas partial wetting (PW) is induced by a negative $A$. 

We distinguish the following regimes (when
$h=0$) \cite{footnote8}: 
\begin{enumerate}

\item strong healing length asymmetry $\xi_{_2}\ll\xi_{_1}$: CW
and $\xi_{_c}=\xi_{_1}/(2\sqrt{K+1})$.

\item strong healing length asymmetry $\xi_{_1}\ll\xi_{_2}$: PW
and $\xi_{_c}=\xi_{_2}/2$.

\item Strong segregation $1/K=0$: PW and
$\xi_{_c}=\xi_{_2}/\sqrt{2}$.

 \item Both strong segregation $1/K\ll 1$
\textit{and} strong healing length asymmetry
$\xi_{_2}\ll\xi_{_1}$:
\begin{enumerate}
\item PW and $\xi_{_c}=\xi_{_2}/\sqrt{2}$ for
$[[\xi_{_2}/\xi_{_1}] \sqrt{K}]^{-1}\lesssim 1$.

\item transition from PW to CW and
$\xi_{_c}=\xi_{_1}/(2\sqrt{K+1})$ for $[[\xi_{_2}/\xi_{_1}]
\sqrt{K}]^{-1}\gtrsim 1$.
\end{enumerate}

\item Weak segregation $0< K-1\ll 1$ and  $0< \xi_{_1}/\xi_{_2}-1\ll 1$:
transition from PW to CW and $\xi_{_c}=\xi_{_2}/\sqrt{K-1}$.

\end{enumerate}
For the cases $4$ and $5$, the transition from PW to CW is
mediated by a completely flat interface potential, that is,
$V(\ell)=0$ for all $\ell$. This observation explains several earlier reported features of the extraordinary wetting (and prewetting) phase transition in this system. Of particular interest is case $5$ for
which we obtained an analytical expression of the full interface potential \eqref{intpot4}, the leading term of which, for large $\ell$, was 
used as a typical model interface potential for calculating the line tension of a three-phase contact line where the 1-2 interface meets the wall. 

The calculation of the line tension represents the most important physical advance reported here. The use of the IDM has allowed us to calculate properties of the inhomogeneous three-phase contact line based on the knowledge of the $V(\ell)$ calculated for homogeneous states. The first of these properties is the interface displacement profile $\ell(x)$ for partial wetting states. Close to wetting these profiles are akin to those normally expected near a critical wetting transition, in spite of the fact that the wetting transition here is of first order. The second property is the line tension $\tau$ for which we have analytic results from the IDM, as well as an analytic conjecture that satisfies all physical requirements for arbitrary contact angle $0<\theta<\pi$ and captures the precise singularity at wetting (or drying). 

The fact that the line tension is maximal at wetting is in accord with the predictions from and expectations raised by other models and theories of the line tension \cite{indekeu4}. However, the fact that $\tau$ approaches zero at wetting (from negative values) and the precise linear dependence on $\theta$ with which it does so, is reminiscent of a mean-field critical wetting transition (for short-range forces) rather than a first-order one. This hybrid character of $\tau$ (reinforced by the fact that the boundary tension along prewetting is zero) is explained by the fully flat barrierless $V(\ell)$ at wetting, which we calculated in this work.

\section{Acknowledgements}
The authors acknowledge partial support by Projects Nos.~FWO
G.0115.06, GOA/2004/02, the KU Leuven Research Fund and useful discussions with Achilleas Lazarides and Dmitry
Tatyanenko, all at the early stages of this research.

\section{Appendix}
What follows are the solutions for the interface potentials as
closed-form integral expressions wherein the parameter $\eta$ must
be eliminated to obtain the relation between $V(\ell)$ and $\ell$.
Note that one may find (rather elaborate) analytic expressions in
terms of hyperbolic integrals for the expressions which follow.
\begin{enumerate}
\item Define first the functional and functions:
\begin{subequations}\label{functies}
\begin{align}
\widehat{\mathcal{Z}}_{_0}(A,\,&B;[C(r),\,D(r)])\equiv\int_{\sqrt{A}}^{\sqrt{B}}
\frac{\text{d}r}{\sqrt{D(r)}}\left(D(r)+C(r)\right),\\
\mathcal{Z}_{_1}(r)&\equiv 1-2\eta r^{2}+r^{4},\\
\mathcal{Z}_{_2}(r)&\equiv 1-\eta^2+2r^{2} (\eta K-1)+(1-K^2)r^{4},\\
\mathcal{Z}_{_3}(r)&\equiv 2r^{2}(K-\eta)+(1-K^2)r^{4},\\
\mathcal{Z}_{_4}(r)&\equiv 1-2 r^{2}+r^{4},\\
\mathcal{Z}_{_5}(r)&\equiv r^2(\eta-1),\\
\mathcal{Z}_{_6}(r)&\equiv r^2-\mathcal{Z}_{_3}(r),\\
\mathcal{Z}_{_7}(r)&\equiv r^2-\mathcal{Z}_{_1}(r),\\
\mathcal{Z}_{_8}(r)&\equiv(\eta-1)(\eta-Kr^2),\\
\mathcal{Z}_{_9}(r)&\equiv\eta-Kr^2-\mathcal{Z}_{_2}(r).
\end{align}
\end{subequations}

\item We briefly explain now how to obtain the interface potential
for the case $\xi_{_2}/\xi_{_1}=0$, as treated in
section~\ref{Asmallcoherencelength}. The potentials of
section~\ref{Alargecoherencelength}
and~\ref{LargeInterPhaseRepulsionK} are given below and can be
obtained in a similar fashion. Looking at Fig.~\ref{fig3}, one
sees that one can split the path of the densities
$(\psi_{_1},\psi_{_2})$ into three parts. Since
$\xi_{_2}/\xi_{_1}=0$, there is no energy contained in the
vertical path with $\psi_{_1}=0$. The equations of motion for the
curved path are given in \eqref{curve} for which the
conservation of energy yields:
\begin{align}\label{COE1}
\xi_{_1}^2\dot{\psi^2_{_1}}&=\frac{1-\eta^2}{2}+\psi_{_1}^2(\eta
K-1)+\frac{1-K^2}{2}\psi_{_1}^4.
\end{align}
Further, the horizontal path starts in ($\sqrt{ K/\eta},\,0$) and
arrives in $(1 ,\, 0)$ as depicted in Fig.~\ref{fig3}. The
equations of motion are there governed by \eqref{horizontal1}
with the following conservation of energy:
\begin{align}\label{COE2}
\xi_{_1}^2\dot{\psi^2_{_1}}=\frac{1}{2}-\psi_{_1}^2+ \psi_{_1}^4.
\end{align}
Writing out expression~\eqref{simpleform2}, we get
\begin{align}\label{simpleform3}
\frac{V(\ell)+\gamma_{_0}-h\ell}{2P_{_1}\xi_{_1}}
&=\int_{_0}^{_\infty}\text{d}z\,\left[2\xi_{_1}^2\dot{\psi_{_1}^2}+
(\eta-1)\psi_{_2}^2\right].
\end{align}
Splitting the integrals into the parts $[0,z_{_0}]$ and
$[z_{_0},\infty]$, using the transformation
$\text{d}z=\text{d}\psi_{_1}/\dot{\psi}_{_1}$, combined with the
conservation laws~\eqref{COE1} and~\eqref{COE2}, we then find
\begin{align}\label{intpot1}
\frac{V(\ell)+\gamma_{_0}-h\ell}{2\sqrt{2}P_{_1}\xi_{_1}}=&
\widehat{\mathcal{Z}}_{_0}\left(\,\eta/K,\,1;[0,\mathcal{Z}_{_4}(r)]\right)\nonumber\\
&+\widehat{\mathcal{Z}}_{_0}\left(0,\,\eta/K;\left[
\mathcal{Z}_{_8},\,\mathcal{Z}_{_2}(r)\right]\right),
\end{align}
with the functions $\mathcal{Z}_{_8}$, $\mathcal{Z}_{_2}$ and
$\mathcal{Z}_{_4}$ defined in~\eqref{functies}. We eliminate the
multiplier $\eta<1$ by writing:
\begin{align}\label{length1}
\frac{\ell}{\xi_{_1}}=\sqrt{2}\widehat{\mathcal{Z}}_{_0}&\left(0,\,\eta/K;
[\mathcal{Z}_{_9}(r),\,\mathcal{Z}_{_2}(r)]\right).
\end{align}
The coefficient $\overline{A}_{_{1}}$ in
\eqref{nogesnenandere} is:
\begin{align}\label{weennul}
&\overline{A}_{_1}=\frac{128\sqrt{2}}{3}
\frac{(2K+1)(K-1)^3}{K^{3/2}(K+1)^2}\\
&\times\left[\frac{\sqrt{2K}+\sqrt{K-1}}{\sqrt{2K}-\sqrt{K-1}}\right]\exp\left[4K\left(1-\sqrt{\frac{K-1}{2K}}\right)\right].\nonumber
\end{align}

 \item For the wall-adsorbed states in the case of $\xi_{_1}/\xi_{_2}=0$,
 the interface potential is found to be a solution of the
following two Eqs. with $\eta>1$:
\begin{subequations}
\begin{align}\label{intpot2Wall}
&\frac{V(\ell)+\gamma_{_0}-h\ell}{2\sqrt{2}P_{_1}\xi_{_2}}=
\widehat{\mathcal{Z}}_{_0}\left(0,\,1/K;[\mathcal{Z}_{_5}(r),\,\mathcal{Z}_{_3}(r)]\right)\\
&\quad\quad\quad\quad+\widehat{\mathcal{Z}}_{_0}\left(0,\,\eta-\sqrt{\eta^2-1};[\mathcal{Z}_{_5}(r) ,\,\mathcal{Z}_{_1}(r)]\right)\nonumber\\
&\quad\quad\quad\quad+\widehat{\mathcal{Z}}_{_0}\left(1/K,\,\eta-\sqrt{\eta^2-1};[\mathcal{Z}_{_5}(r)
,\,\mathcal{Z}_{_1}(r)]\right),\nonumber\\
&\frac{\ell}{\xi_{_2}}=\sqrt{2}\,\widehat{\mathcal{Z}}_{_0}\left(0,\,1/K;[\mathcal{Z}_{_6}(r),\,\mathcal{Z}_{_3}(r)]\right)\\
&\quad\quad\quad+\sqrt{2}\,\widehat{\mathcal{Z}}_{_0}\left(1/K,\,\eta-\sqrt{\eta^2-1};[\mathcal{Z}_{_7}(r),\,\mathcal{Z}_{_1}(r)]\right)\nonumber\\
&\quad\quad\quad+\sqrt{2}\,
\widehat{\mathcal{Z}}_{_0}\left(0,\,\eta-\sqrt{\eta^2-1};[\mathcal{Z}_{_7}(r)
,\,\mathcal{Z}_{_1}(r)]\right).\nonumber
\end{align}
\end{subequations}

 \item For the bulk states with $\xi_{_1}/\xi_{_2}=0$, we must distinguish two cases.
First of all, when the multiplier $\eta$ takes the values
$(1+K^2)/(2K)<\eta<K$, we have:
\begin{subequations}
\begin{align}\label{intpot2bulk1}
&\frac{V(\ell)+\gamma_{_0}-h\ell}{4\sqrt{2}P_{_1}\xi_{_2}}=
\,\widehat{\mathcal{Z}}_{_0}\left(0,\,\frac{2(K-\eta)}{K^2-1};[\mathcal{Z}_{_5}(r)
,\,\mathcal{Z}_{_3}(r)]\right),\\
&\frac{\ell}{\xi_{_2}}=2\sqrt{2}\,\widehat{\mathcal{Z}}_{_0}\left(0,\,\frac{2(K-\eta)}{K^2-1};[\mathcal{Z}_{_6}(r)
,\,\mathcal{Z}_{_3}(r)]\right).
\end{align}
\end{subequations}
 Secondly, when $1<\eta<(1+K^2)/(2K)$, we have:
\begin{subequations}
\begin{align}\label{intpot2bulk2}
&\frac{V(\ell)+\gamma_{_0}-h\ell}{4\sqrt{2}P_{_1}\xi_{_2}}=
\,\widehat{\mathcal{Z}}_{_0}\left(0,\,1/K;[\mathcal{Z}_{_5}(r)
,\,\mathcal{Z}_{_3}(r)]\right)\\
&+\,\widehat{\mathcal{Z}}_{_0}\left(1/K,\,\eta-\sqrt{\eta^2-1};[\mathcal{Z}_{_5}(r),\,\mathcal{Z}_{_1}(r)]\right),\nonumber\\
&\frac{\ell}{\xi_{_2}}=2\sqrt{2}\,\widehat{\mathcal{Z}}_{_0}\left(0,\,1/K;[\mathcal{Z}_{_6}(r),\,\mathcal{Z}_{_3}(r)]\right)\\
&+2\sqrt{2}\,\widehat{\mathcal{Z}}_{_0}\left(1/K,\,\eta-\sqrt{\eta^2-1};[\mathcal{Z}_{_7}(r),\,\mathcal{Z}_{_1}(r)]\right).\nonumber
\end{align}
\end{subequations}

\item When $1/K=0$, the interface potential is found through a solution
of the following two Eqs. with $\eta>1$:
\begin{subequations}
\begin{align}\label{intpot3}
&\frac{V(\ell)+\gamma_{_0}-h\ell}{4\sqrt{2}P_{_1}\xi_{_2}}
 =\,\widehat{\mathcal{Z}}_{_0}\left(0,\eta-\sqrt{\eta^2-1};[\mathcal{Z}_{_5}(r),\,\mathcal{Z}_{_1}(r)]\right),\\
&\frac{\ell}{\xi_{_2}}=2\sqrt{2}
\widehat{\mathcal{Z}}_{_0}\left(0,\eta-\sqrt{\eta^2-1};[\mathcal{Z}_{_6}(r),\,\mathcal{Z}_{_1}(r)]\right).
\end{align}
\end{subequations}
 \item The exact interface potential in the case
 $0< K-1\ll 1$ and $0< \xi_{_1}/\xi_{_2}-1\ll 1$ is
\begin{align}\label{intpot4}
&V(\ell)=\frac{1}{\sqrt{2}}\left[\frac{2\sqrt{2}(\xi_{_1}/\xi_{_2}-1)}{3\sqrt{K-1}}-1\right]\times\\
&\text{LambertW}\left(-2\sqrt{2}\sqrt{K-1}\,
e^{-2\sqrt{K-1}\,(\ell/\xi_{_2}+\sqrt{2})}\right),\nonumber
\end{align}
where LambertW$(x)$ is the solution $y(x)$ to the equation
$y(x)e^{y(x)}=x$.
\end{enumerate}


\begin{thebibliography}{99}
\bibitem{footnote1}Three
remarks can be made. 1) The system can also be seen as a
one-particle system in two dimensions in which the particle has
anisotropic mass. 2) Rotational symmetry of the potential is only recovered
for $K=1$. 3) Minima of the excess energy $\gamma$
correspond to maxima of the potential $U$.
\bibitem{footnote2} In what follows we draw
$(\psi_{_1},\psi_{_2})$-paths, rather than interface density
profiles. Thereby, we gain the possibility of showing the
potential $U$ in which the particles move, together with the
influence of the parameters $K$ and $\xi_{_2}/\xi_{_1}$.  Showing the $(\psi_{_1},\psi_{_2})$-paths is particularly useful when the system involves two largely different
length scales (being
the time scales in the dynamical approach). In that case, spatial density profiles are difficult to visualize.
\bibitem{footnote3} Therefore also, the healing length of
species $2$ equals $\xi_{_2}\equiv \hslash/\sqrt{2m_{_2}\overline{\mu}_{_2}}$
instead of $\xi_{_2}\equiv\hslash/\sqrt{2m_{_2}\mu_{_2}}$.
\bibitem{footnote4} The fact that length $\xi_{_c}$ must be finite
when $K=1$ and $\xi_{_1}\ll\xi_{_2}$, and therefore $\xi_{_c}\neq
\Lambda_{_2}$, can be understood as follows. If indeed $\xi_{_c}$
were to diverge at $K=1$, for nonvanishing coefficients in front
of the exponents, this would give rise to a constant interface
potential. However, this is incompatible with the nonzero
spreading coefficient.
\bibitem{footnote5} We were unable to prove this statement analytically. The
difficulty is that, for a fixed value of the film thickness, the
corresponding disjoining pressures of bulk-nucleated and
wall-adsorbed states differ.
\bibitem{footnote6} In fact, we first consider
$1/K$ to be very small but nonzero so that some overlap between
species $1$ and $2$ persists and thus
expression~\eqref{consenergy} is still valid. Only then we take
the limit $1/K \rightarrow 0$.
\bibitem{footnote7}It is unclear at
this point whether the deviations around $\kappa=3/\sqrt{2}$ as
seen in Fig.~\ref{fig11} are due to numerical errors or not.
\bibitem{footnote8} Within the domain where the GP formalism is, to a
good approximation, valid, the here obtained interface potential
will remain as it is. Indeed, for small but nonzero absolute temperature $T$
and upon ignoring quantum fluctuations, thermal fluctuations will
renormalize the interface potential; it is known that the
corrections can be developed as a function of the parameter
$\omega=k_{_B}T/(\gamma_{_{12}} \xi_{_c}^2)$ where $\xi_{_c}$ is
the decay length of the interface potential; a
fluctuation-dominated regime arises when the parameter $\omega$ is
of order unity or larger. For the system under consideration we find
$\omega\ll 1$. Indeed, since $\xi_{_c}$ is always larger than the
healing length $\xi_{_1}$ (take for simplicity that
$\xi_{_1}\approx\xi_{_2}$), one finds that
$\omega<k_{_B}T/(\mu_{_1} (\xi_{_1}/a_{_1}))$. However, we know
that in the current experiments $k_{_B}T<\mu_{_1}$ and
$\xi_{_1}/a_{_1}\gg 1$ such that $\omega\ll 1$~\cite{pita}.
Therefore, at low temperature, thermal fluctuations of the
interface are expected to alter the interface potentials which
were obtained in this chapter, only weakly. As was proven in
\cite{vanschaeybroeck}, the finite-temperature
corrections to the interfacial tension are small, except possibly for
weak segregation.
\end{thebibliography}
\end{document}